\documentstyle[color,amsfonts,amsmath,amssymb,mathrsfs,verbatim,11pt,epsf]{article}
\def\inn{\in}

\def\sigsum{\mbox{{\Large $\sum$}$ \!\!\!\!\!\!  {\raisebox{-2.4ex}{\makebox[1em]{\scriptsize $r\neq %%@
1,q$}}} \; $}}
\def\pqg{\mbox{$ \!\! \mbox{\Large \_} \! $}}
\def\pqo{\mbox{$ \!\!\!\!\; \mbox{\textbf {\Large \_}} \! $}}
\def\sp{\; \!}
\def\sb{\;\;}
\def\kl{k \pqo l}
\def\jl{j \pqo l}

\def\il{i \mbox{$ \!\!\!\!\; \:\! \mbox{\textbf{\Large \_}} \! \!\!\; $} l}
\def\scirc{\mbox{\raisebox{0.2ex}{\scriptsize {$\circ$}}}}

\def\TM{\mbox{\it TM}}

\def\Sbar{S\mbox{$\!\!\!\!\!\: \mbox{{\raisebox{0.3ex}{\small{--}\!\!\small{--}\!\!}} }$}}
\def\Sbard{\dot{S}\mbox{$\!\!\!\!\!\: \mbox{{\raisebox{0.3ex}{\small{--}\!\!\small{--}\!\!}} }$}}

\def\rrr{{\mathbb{R}}}
\def\ccc{{\mathbb{C}}}
\def\hhh{{\mathbb{H}}}
\def\ooo{{\mathbb{O}}}

\def\b1{\mbox{\boldmath $1$}}
\def\v{\mbox{\boldmath $v$}}

\def\bh{\mbox{\boldmath $h$}}

\def\br{\mbox{\boldmath $r$}}
\def\bv{\mbox{\boldmath $v$}}
\def\bvh{\hat{\bv}}

\def\bu{\mbox{\boldmath $u$}}

\def\bsig{\boldsymbol {\sigma}}

\def\lv{L(\v)=1}
\def\lvh{L(\hat{\v})=1}

\def\lvf{L(\v_4)=1}

\def\lvte{L(\v_{10})=1}
\def\lvt{L(\v_{27})=1}
\def\lvfs{L(\v_{56})=1}
\def\lvtfe{L(\v_{248})=1}

\def\htwc{\mbox{h}_2\ccc}
\def\hthc{\mbox{h}_3\ccc}

\def\htwo{\mbox{h}_2\ooo}
\def\htho{\mbox{h}_3\ooo}

\def\sltc{\mbox{SL}(2,\ccc)}
\def\sltca{\mbox{sl}(2,\ccc)}

\def\slthc{\mbox{SL}(3,\ccc)}

\def\sltho{\mbox{SL}(3,\ooo)}

\def\sltwoo{\mbox{SL}(2,\ooo)}

\def\slthoa{\mbox{sl}(3,\ooo)}

\def\uo{\mbox{U}(1)}

\def\sutw{\mbox{SU}(2)}

\def\suth{\mbox{SU}(3)}
\def\sutha{\mbox{su}(3)}
\def\spot{\mbox{Spin}^+(1,3)}
\def\spotn{\mbox{Spin}^+(1,9)}

\def\soot{\mbox{SO}^+(1,3)}

\def\sootn{\mbox{SO}^+(1,9)}

\def\gt{\mbox{G}_2}
\def\ff{\mbox{F}_4}
\def\ee{\mbox{E}_8}
\def\eeg{\mbox{E}_{8(-24)}}
\def\ese{\mbox{E}_7}
\def\eseg{\mbox{E}_{7(-25)}}
\def\esi{\mbox{E}_6}
\def\esig{\mbox{E}_{6(-26)}}

\def\SML{\mbox{SU}(3)_c \times \mbox{SU}(2)_L \times \mbox{U}(1)_Y}
\def\stab{\mbox{Stab}(\TM_4)}
\def\stabse{\mbox{Stab}_7(\TM_4)}

\def\staba{\mbox{stab}(\TM_4)}

\def\thetmt{\theta_{\! M^2}}

\def\thX{{\theta^{1}_{\! \mbox{\tiny{$X$}}}}}
\def\thY{{\theta^{1}_{\! \mbox{\tiny{$Y$}}}}}

\def\lag{{\mathcal L}}

\def\mcM{{\mathcal M}}

\def\mcX{{\mathcal X}}
\def\mcY{{\mathcal Y}}

\def\fh{\frac{1}{2}}
\def\fhs{\mbox{\small{$\fh$}}}

\def\ol{\overline}
\def\ul{\underline}

\def\gpath{ }
\def\maxwidth{14.4cm}

\def\setb{\setlength{\baselineskip}{0.625\baselineskip}}

\begin{document} 

% take out double spacing (28/11/15)
{\setlength{\baselineskip}{0.625\baselineskip}

%\begin{titlepage}
%\bigskip

\begin{center}
%\bigskip
 
 {\LARGE{\bf  A Novel Approach to Extra Dimensions }} \\

\bigskip
%\bigskip
%\vspace{40pt}

\mbox {{\Large David J. Jackson} }  \\

%  \vspace{20pt}

% { \large }  
  
  \vspace{10pt}
 
%\today
 { \large March 2, 2016 }

 \vspace{10pt}

{\bf  Abstract}

\end{center}

   Four-dimensional spacetime, together with a natural generalisation to extra dimensions, is obtained %%@
through an analysis of the structures and symmetries deriving from possible arithmetic expressions for %%@
one-dimensional time.  
  On taking the infinitesimal limit this simple one-dimensional structure can be consistently equated with %%@
a homogeneous form of arbitrary dimension possessing both spacetime and more general symmetries. An %%@
extended 4-dimensional manifold, with the associated spacetime symmetry,  provides a natural breaking %%@
mechanism for a higher-dimensional form and symmetry of time. It will be described how this symmetry %%@
breaking leads to a series of distinct properties of the Standard Model of particle physics, deriving %%@
directly from the natural mathematical development of the theory.
 
\vspace{10pt}

%\tableofcontents

%\pagebreak

{ %fit table of contents onto two pages
\setlength{\baselineskip}{0.5\baselineskip}
%\setb
\parskip -0.05pt plus 1pt minus 1pt

\tableofcontents

}

%\end{titlepage}

%\clearpage

%\pagebreak

%\pagebreak

\section{Introduction}

  In 4-dimensional spacetime an interval of proper time $ds$ can be expressed locally %%@
in terms of the Minkowski metric $\eta = \mbox{diag}(+1,-1,-1,-1)$ for suitable local %%@
coordinates $x^a$ as:  
\begin{equation}
    ds^2 = \eta_{ab}dx^a dx^b
	\label{dsxx}
\end{equation}
 with $a,b \in \{0,1,2,3\}$.
  A typical approach to extra dimensions would involve an extension to the range of %%@
the indices $a,b$ in the above quadratic expression, with the metric correspondingly %%@
replaced by an $n \times n$ matrix for the $n$-dimensional spacetime generalisation. %%@
An extended 4-dimensional spacetime manifold, with the local metric of %%@
equation~\ref{dsxx}, might be identified for example through the spontaneous %%@
compactification of the extra dimensions, in principle resulting in residual physical %%@
properties that might be observable in 4-dimensional spacetime.

  For the theory described in this paper in place of generalising  the right-hand side %%@
of equation~\ref{dsxx} for a higher-dimensional \textit{spacetime} structure we %%@
consider a generalisation of the overall expression as constrained by the %%@
\textit{time} interval on the left-hand side. For example from the perspective of the %%@
linear one-dimensional flow of time it is equally permitted to write the cubic %%@
expression:
\begin{equation}
    ds^3 = \alpha_{abc}dx^a dx^b dx^c
	\label{dsxxx}
\end{equation} 
   with each coefficient $\alpha_{abc} \in \{-1,0,1\}$ for $a,b,c \in \{1\ldots n\}$ %%@
for the $n$-dimensional case. In the following section we describe how to express the %%@
general form of this type of generalisation. The means of identifying the %%@
4-dimensional structure of equation~\ref{dsxx} within this generalisation will be %%@
described in section~\ref{chaputtf} and will provide the means of breaking the full %%@
symmetry of expressions such as equation~\ref{dsxxx}.

   In sections~\ref{esihtho}--\ref{secesef} it will be described how natural %%@
mathematical extensions of this idea lead directly to the identification of physical %%@
properties such as fractional charges for quark states and a left-right asymmetry of a %%@
kind closely resembling the structure of the Standard Model. This development leads to %%@
the consideration of an $\ee$ symmetry in section~\ref{sosmfi} as a final stage in %%@
this progression that might in principle accommodate the full set of Standard Model %%@
properties.

\section{Time and Spacetime}
\label{sym}

   In this paper it is described how consideration of a multi-dimensional form of time 
and symmetry breaking over a 4-dimensional spacetime manifold leads directly to %%@
structures which exhibit a close resemblance to the Standard Model. 
   We begin here by developing  the underlying idea.
     A finite interval of time represented by the real number $s\inn \rrr$ can be %%@
algebraically expressed in terms of other real numbers $x^a$ $(a=1,2,3\ldots)$ in an %%@
endless variety of ways, for example we can have $s=x^1(x^2x^3+x^4)$ and so on simply %%@
by employing the basic arithmetic structure of the real line.

    The broad range of possible expressions for a finite interval $s$ in terms of an %%@
arbitrary number of variables \{$x^a$\}, $a = 1\ldots n$, will be constrained to a %%@
more  
 restrictive structure  in the limit of infinitesimally small temporal intervals. We %%@
first consider this limit for the trivial case with the flow of time $s$ expressed in %%@
terms of a single real variable $x^1$ only for which we have simply $s=x^1$. This can %%@
symbolically be written as $\delta s=\delta x^1$ as we approach the limit of %%@
infinitesimal intervals. We then express the rate of change of $x$ with respect to $s$ %%@
in this limit as:   
\begin{equation}
   \label{vonedo}
   v^1= \frac{dx^1}{ds} \equiv \frac{\delta x^1}{\delta s}\bigg{\vert}_{\delta s \to %%@
0}  = 1 	   
\end{equation}
   For the case with multiple real numbers $\{x^a\}\inn\rrr^n$ representing the flow %%@
of time $s$ each will be associated with a corresponding rate of change $v^a=dx^a/ds$ %%@
with respect to  time. For example, we may consider the propagation of time expressed %%@
for an infinitesimal interval as:
\begin{eqnarray}
 (\delta s)^2 & = & (\delta x^1)^2 + (\delta x^2)^2 + (\delta x^3)^2  \label{flow3del} %%@
\\
         & = & \eta_{ab}  \delta x^a\delta x^b  \qquad \mbox{with}  \quad
		         \eta = \mbox{diag}(+1,+1,+1)  \label{etamet}
\end{eqnarray}
  where $a,b \in \{1,2,3\}$ (and with the conventional summation over repeated indices %%@
implied throughout this paper). Dividing by $(\delta s)^2$ and taking the limit %%@
$\delta s \to 0$ this can be written as $\eta_{ab}v^a v^b=1$ or %%@
$(v^1)^2+(v^2)^2+(v^3)^2=1$, which is invariant under the group, O(3), of orthogonal %%@
transformations in three dimensions applied to $\bv_3 = (v^1,v^2,v^3) \inn \rrr^3$.
 The question is then how to express the general case for the composition and %%@
symmetries of a multi-dimensional set of velocities $\{v^a\} \inn \rrr^n$.

     The infinitesimal elements of time can be written most generally, taking care to %%@
balance the order of the vanishing elements in each term, as:
\begin{equation}
\label{sbits}
  \delta s = \alpha_a\delta x^a + \sqrt{\alpha_{bc}\delta x^b \delta x^c} +
  \;\,  {}^{\substack{ 3 \vspace{0.6mm} \\ {} }} \!\!\!\!\sqrt{\alpha_{def}\delta x^d %%@
\delta x^e \delta x^f} + \ldots\ldots
\end{equation}
                   Here the coefficients $\alpha_{abc\ldots}$ are each equal to $\pm %%@
1$ or $0$ since we wish to express the $\delta s$ purely in terms of simple arithmetic %%@
relations of the $\delta x^a$. In equation~\ref{sbits} each term divides $\delta s$ %%@
into a separate portion of time:
\begin{equation}
  \delta s \quad =\quad \delta s_1 \qquad +\qquad \delta s_2
   \qquad +\qquad \delta s_3 \qquad+\qquad \ldots
\end{equation}
                   where each term $\delta s_p$ is the $p^{\mathrm{th}}$-root of a %%@
homogeneous polynomial of order $p$ in the \{$\delta x^a$\}. Taking each term in turn, %%@
dividing by the interval $\delta s_p$ in each case and taking the limit $\{ \delta s_p %%@
, \delta x^a \} \to 0$ we find:
\begin{eqnarray}
         \delta s_p & = & \;\,  {}^{\substack{ p \vspace{0.6mm} \\ {} }} \!\!\!\! %%@
\sqrt{\alpha_{abc\ldots}\delta x^a \delta x^b \delta x^c\ldots} \\   
    \mbox{divide by $\delta s_p$:$\;\;$} \qquad \quad \qquad \qquad   1 & = & \;\,  %%@
{}^{\substack{ p \vspace{0.6mm} \\ {} }} \!\!\!\! \sqrt{\alpha_{abc\ldots}v^a v^b %%@
v^c\ldots} \label{pone} \\ 
             \mbox{that is:}   \qquad    \alpha_{abc\ldots}v^a v^b v^c\ldots  & = & 1  %%@
\label{ptwo} \\ 
         \mbox{which we write:} \qquad  \qquad \qquad    L(\v) & = & 1   \label{lv}
\end{eqnarray}      
                   where $L$ is a homogeneous polynomial of order $p$ in the %%@
components $v^a$; it can be considered as a map from the elements of a real %%@
$n$-dimensional vector space $\v \inn \rrr^n$ onto the unit $1\inn\rrr$.
Equation~\ref{lv} is taken to express the general mathematical form of %%@
multi-dimensional temporal flow and it is the central equation of this paper.   
  The symmetries of $L(\v)= 1$ will be represented by groups acting on the vector %%@
space $\rrr^n$ such that for all elements $g$ of the group $G$ and all vectors $\v\inn %%@
\rrr^n$ satisfying $\lv$ we have $L(\sigma_g(\v)) = L(\v') = 1$ where $\sigma_g(\v)$ %%@
represents the action of the group element $g \inn G$ on the vector $\v\inn \rrr^n$.

     Quadratic forms in general, including the 4-dimensional  form:
\begin{equation}
L(\bv_4)  = (v^0)^2- (v^1)^2-(v^2)^2-(v^3)^2 = 1  \label{flow4d}    
\end{equation}    
   that is $L(\bv_4) = \eta_{ab}v^av^b = 1$ for $a,b \in \{0,1,2,3\}$ with the %%@
Minkowski metric $\eta = \mbox{diag}(+1,-1,-1,-1)$, $\bv_4 \inn \rrr^4$ and with %%@
Lorentz $\soot$ symmetry,	  and the norm of an element of a division algebra %%@
($\rrr,\ccc,\hhh$ or $\ooo$), together with their symmetry groups, are expected to be %%@
particularly significant forms of $\lv$. This is due to their close relation to %%@
Clifford algebras and Euclidean spatial geometry, describing the geometry of external %%@
space and spacetime.

  Equation~\ref{flow4d} is the special case of equation~\ref{ptwo} in the form of the %%@
particular case of equation~\ref{dsxx} described in the introduction. The %%@
generalisation to the cubic form in equation~\ref{dsxxx} is incorporated within the %%@
general expressions of equations~\ref{ptwo} and \ref{lv}.
  Other possible forms of $\lv$ include the determinants of matrices, which are %%@
homogeneous polynomials in the matrix elements.

 With various different forms of progression in time to be considered, in general the %%@
subscript $n$ in the notation $L(\bv_n)=1$ indicates collectively the vector space %%@
$\rrr^n$, the implied form $L$ and the corresponding symmetry group $G$ (respectively %%@
$\bv_{4} \inn  \rrr^{4}$, $L(\bv_4) = \eta_{ab}v^av^b = 1$ and $G=\soot$ in the above %%@
example for $n=4$). The notation $\lvh$ and $\hat{G}$, with a `hat' above a kernel %%@
symbol, will denote the highest-dimensional form of time considered and its symmetry %%@
respectively.

   Given a possible $n$-dimensional form of progression in time, $L(\bv_n)=1$, the %%@
vector $\bv_n \inn \rrr^n$ may be written as the ordered set of velocities: 
\begin{eqnarray}
      \bv_n & = & \, \{ \: v^1, \quad  v^2,\ldots \quad v^n \, \}  \label{vnset}  \\   
            & = &  \bigg\{ \frac{dx^1}{ds},  \frac{dx^2}{ds},\ldots  \frac{dx^n}{ds} %%@
\bigg\}  
\end{eqnarray}  
    the values of which are unchanged by a numerical translation of the real %%@
variables,
\begin{equation}
    x^a \to x^a + r^a
\end{equation}
    for any constant set $\{ r^a \} = \br_{\! n} \inn \rrr^n$, or for a subset of %%@
$\rrr^n$. 
	Above we described a possible symmetry of $\lv$ with the action of a group $G$ %%@
mixing the numerical components $v^a$, which represent elements of the flow of time  %%@
$dx^a/ds$. Here we have a further symmetry implicit in $\lv$ with respect to %%@
translations of the numerical variables as $x^a \to x^a + r^a$.
	That is, we also have trivially:
\begin{eqnarray}
  \label{rspill}
      \bv_n & = &  \bigg\{\frac{d(x^1+r^1)}{ds},\frac{d(x^2+r^2)}{ds},\ldots  %%@
\frac{d(x^n+r^n)}{ds}\bigg\}.  
\end{eqnarray}  
  satisfying $L(\bv_n) = 1$.
The relation between the `translation symmetry' of $\lv$ and the `rotation symmetry', %%@
more generally denoted by the action $\sigma_g(\bv)$ for $g\inn G$, is key to the %%@
development of the geometrical structure of the theory.

\section{Extra Dimensions}
\label{chaputtf}

   We initially  consider
  $\hat{G}=\sootn$ to be provisionally taken as the full symmetry group for the form %%@
$L(\bvh)=\lvte$, which in turn is the 10-dimensional extension of %%@
equation~\ref{flow4d} (this is also equivalent to an extension for extra spacetime %%@
dimensions as described for equation~\ref{dsxx}). Here an extended base manifold $M_4$ %%@
arises through employing for four of the ten translational degrees of freedom of %%@
$\lvte$, in the manner described in equation~\ref{rspill}.
 For this model we hence obtain the structures described in figure~\ref{mtogmaphr}.

\vspace{5pt}
\begin{figure}[htbp]  
\centering
\epsfxsize=\maxwidth
\leavevmode
\epsffile[0 0 1882 964]{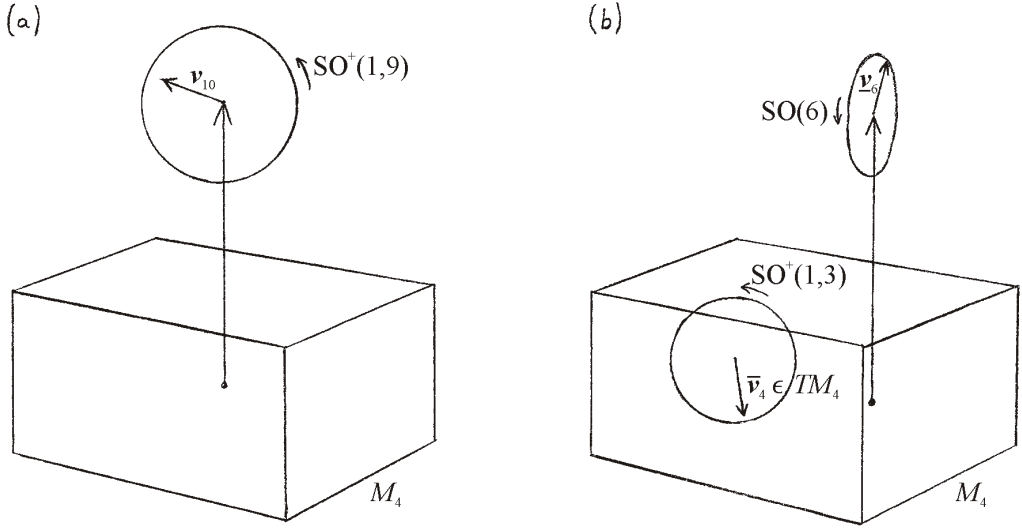}
\vspace{-30pt}
\caption{\setb (a) The full symmetry group $\hat{G}=\sootn$ over the base space $M_4$
(b) broken to the internal symmetry SO(6) with external subgroup $\soot \subset %%@
\sootn$
 acting on the tangent space $\TM_4$. (An underline, as for $\ul{\bv}_6$, or an %%@
overline, as for $\ol{\bv}_4$, may be used for internal and external objects %%@
respectively, in case of ambiguity). }
\label{mtogmaphr}
\end{figure} 

 This figure also represents the manner in which the identification of an extended %%@
4-dimensional background manifold breaks the symmetry of the higher-dimensional form %%@
of time $\lvh$. However while quadratic spacetime forms such as $\lvte$ are included %%@
as a possible structure of equation~\ref{lv} more generally cubic or higher-order %%@
polynomial forms of time are also permitted (as initially described for %%@
equation~\ref{dsxxx}).

  We approach this generalisation via the group $\sltc$ as the two-to-one cover of %%@
$\soot$, which may be exhibited by mapping a Lorentz vector $\bv_4 \inn \rrr^{1,3}$ %%@
into the space of $2\times 2$ complex Hermitian matrices as:
\begin{equation}
\label{vtoh}
  \v_4 = (v^0,v^1,v^2,v^3)\; \to\; \bh_2 = \bv_4 \! \cdot \! \bsig =
   \left( \begin{array}{cc} v^0+v^3 & v^1-v^2i \\ v^1+v^2i & v^0-v^3 \end{array}  %%@
\right) \subset \htwc
\end{equation}
 where $\bsig$ denotes the $2\times 2$ identity matrix $\sigma^0$ together with the %%@
three Pauli matrices $\sigma^a$, that is $\sigma^1 = \binom{0 \; 1}{1 \; 0}$,  %%@
$\sigma^2 = \binom{0 \, -i}{i \,\; 0}$, $\sigma^3 = \binom{1 \,\; 0}{0 \, -1}$.
 While the fundamental representation of $\sltc$ acts on the space $\ccc^2$, the group %%@
action for elements $S\inn \sltc$ on the space $\htwc$ provides another representation %%@
given by: 
\begin{equation}
 \label{hshs}
  \bh_2 \to  \bh_2^{\prime} =  S \, \bh_2 \, S^{\dagger}
\end{equation} 
  This maps $\bh_2 \to \bh_2^{\prime}$ onto a new $2\times 2$ complex Hermitian matrix %%@
with the same determinant; hence mapping the components $v^a_4 \to v^{\prime a}_4$ %%@
according to a Lorentz transformation of the real 4-vector $\v_4\inn\rrr^{1,3}$.
 This $\sltc$ action expresses the symmetry of $\lvf$ of equation~\ref{flow4d} in a %%@
manner that naturally extends, consistent with equation~\ref{lv}, to an $\slthc$ %%@
symmetry of the cubic polynomial form $L(\bv_9) = \det(\bv_9) = 1$ with $\bv_9  \inn %%@
\hthc$. For this latter case, in identifying the base space $M_4$ through the %%@
translation symmetry of subspace of external vectors $\ol{\bv}_4 \equiv \bh_2 \inn %%@
\htwc \subset \hthc$, the symmetry is broken to $\sltc \times \uo \subset \slthc$.    
  The action of the external symmetry
 $\sltc$ on the full space $\hthc$ may then be considered. The $2 \times 2$ matrices %%@
$S\inn \sltc$ can be embedded in $3 \times 3$ matrices acting on $\bv_9 \inn \hthc$ %%@
as:
\begin{equation}
     \label{vinhinc}
          \bv_9 \;\, \to \;\,
		\left( \begin{array}{c|c} 
        \,\,\,\,\:\! S \;\!\!   \begin{array}{cc} &  \\  &  \end{array} \!\!\!   &
        \,     0  \begin{array}{cc} &  \\  &  \end{array} \!\!\!\!\!\!\!\!\!\! 
				                         \\  \hline
        \,\,\,\,\,\,  0  \!\! \begin{array}{cc}        &   \end{array}   &	  
		\,  1      \end{array}  \right) 
		\left( \begin{array}{c|c} 
        \,\,\,\, \bh_2     \!\!     \begin{array}{cc} &  \\  &  \end{array} \!\!\!   &
        \,  \psi_L  \begin{array}{cc} &  \\  &  \end{array} \!\!\!\!\!\!\!\!\!\! 
				                         \\  \hline
        \,\,\,\,\,\, \psi_L^{\dagger} \!\!\! \begin{array}{cc}    &   \end{array}   &	  
		\,  n      \end{array}  \right) 
		\left( \begin{array}{c|c} 
        \,\,\,\,\, S^{\dag} \!  \begin{array}{cc} &  \\  &  \end{array} \!\!\!   &
        \,  0  \begin{array}{cc} &  \\  &  \end{array} \!\!\!\!\!\!\!\!\!\! 
				                         \\  \hline
        \,\,\,\,\,\, 0 \!\! \begin{array}{cc}        &   \end{array}   &	  
		\,  1      \end{array}  \right) 
\end{equation}

   This combines the vector representation of $\sltc$ on $\bh_2 \inn \htwc$ and the %%@
spinor representation on $\psi_L \inn \ccc^2$ (taken to be left-handed), together with %%@
the scalar denoted $n \inn \rrr$ (in line with the notation used for %%@
equation~\ref{xoct3} below), in a single symmetry transformation which preserves  %%@
$L(\bv_9) = \det(\bv_9) = 1$.

  Hence by considering a higher-dimensional \textit{cubic} form of time we have %%@
identified components of the `extra dimensions' transforming as a spinor, namely the %%@
$\psi_L$ of equation~\ref{vinhinc}, which also transforms non-trivially under the %%@
internal $\uo$ action identified in the symmetry breaking. The is in contrast to %%@
Kaluza-Klein theories in which a 4-vector object $A_{\mu}$, exhibiting properties of %%@
gauge field, can be identified in the extra components of a higher-dimensional metric %%@
tensor, as for example in the $5 \times 5$ metric case as originally formulated in the %%@
1920s~\cite{Kaluza,Klein}. In the following section we consider a further natural %%@
extension of equation~\ref{vinhinc} for a higher-dimensional form of $\lv$ for the %%@
present theory.

\section{$\esi$ Symmetry on $\htho$}
\label{esihtho}

  A  natural generalisation from the space $\hthc$, underlying the vector $\bv_9 \inn %%@
\hthc$ transformed in equation~\ref{vinhinc} as a symmetry time, is obtained by %%@
augmenting the complex numbers $\ccc$ to the largest division algebra, namely the %%@
octonions $\ooo$~\cite{Baez1}. 
The vector space $\htho$ obtained corresponds to the set of $3\times 3$ Hermitian %%@
matrices over the octonions with elements which can be 
 written as (in this paper we closely follow \cite{Man2}, \cite{Wang} chapters 3 and %%@
4, together with \cite{Man4,Man5,Wang2}, for all details of the $\esi$ structure, and %%@
generally adopt the notation therein):

\begin{equation}
  \label{xoct3}
  {\mathcal X} \; = \;     
	\left( \!\!\!\!\!\!\!\!\;\! \begin{array}{cc} 
               \begin{array}{cc}   p \,\, & \; \bar{a}  \\ a \,\, & \; m \end{array} %%@
\!\!\!\!\!\!   &
          \!     \begin{array}{c}    c  \\  \bar{b}   \end{array} \!\! 
				                         \\  
        \;\;\,\,\,\; \bar{c} \;\;\;\;\;\:\, b \!\!\!\!\!\!\!  \begin{array}{cc}        %%@
&   \end{array}  &	  
		  \:  n      \end{array}  \right) \; = \;
	\left( \begin{array}{c|c} 
        \,\,\,\, X                \begin{array}{cc} &  \\  &  \end{array} \!\!\!   &
        \,  \theta  \begin{array}{cc} &  \\  &  \end{array} \!\!\!\!\!\!\!\!\!\! 
				                         \\  \hline
        \,\,\,\,\,\, \theta^{\dagger} \!\! \begin{array}{cc}        &   \end{array}   %%@
&	  
		\,  n      \end{array}  \right)   
	\;	 \inn \htho  
\end{equation}
   with $p,m,n\inn\rrr$ (here the component labels are chosen to conform with the %%@
notation in the main references, and $n$ here is of course not the dimension of any %%@
space),
 $a,b,c\inn\ooo$  and $\bar{a}$ denotes the octonion conjugate of $a$ reversing the %%@
sign of the 7-dimensional imaginary part. 
  In general an octonion can be described by eight real parameters $\{a_1 \ldots %%@
a_8\}$ and  written as:
\begin{equation}
\label{octa}
  a \; = \; a_1 \; + \; a_2\,i \; + \; a_3\,j + \; a_4\,k \; + \; a_5\,{\kl}
        \; + \; a_6\,{\jl}   \; + \; a_7\,{\il} \; + \; a_8\,l
\end{equation}  
  The seven imaginary units in this basis $\{i,j,k,\kl,\jl,\il,l\}$, with %%@
$i^2=j^2=\ldots=\il^2=l^2 = -1$, are mutually anticommuting, with $\il\sp j = - j %%@
\sp\il$ etc., with their full algebraic composition described in~\cite{Wang,Man4}.
 While the octonions are non-associative they compose the largest `normed division %%@
algebra'~\cite{Baez1}. 
 In equation~\ref{xoct3} $X$ and $\theta$ have the structure of an octonionic $2 %%@
\times 2$ vector and $1 \times 2$ spinor respectively.
        Hence the vector space $\htho$ is 27-dimensional over the real numbers. It is %%@
a space with particularly rich symmetry properties largely owing to the nature of the %%@
8-dimensional octonion subspaces~\cite{Baez1}.

  As for the space $\hthc$ 
 a cubic norm, or determinant $\mbox{det}(\mcX)$ for $\mcX\inn\htho$,
  can be defined on the space $\htho$ taking the form:
\begin{eqnarray}
  \det({\mathcal X})  & = & \det(X)n + 2X\cdot (\theta\theta^{\dag})   \label{detx3}  %%@
\\
                      & = & pmn - p\vert b \vert^2 - m\vert c \vert^2 - n\vert a %%@
\vert^2  
                            + 2 \mbox{Re}(\bar{a}\bar{b}\bar{c})     \label{detpmn}
\end{eqnarray}
   where the 10-dimensional Lorentz inner product $X\cdot Y$  with $X,Y \inn \htwo$, %%@
in the first line, together with equation~\ref{xoct3} can be used to derive the second %%@
line in which the cubic composition of components, consistent with the homogeneous %%@
form of equation~\ref{lv} (and with equation~\ref{dsxxx} as a particular cubic form), %%@
is explicitly seen.
 The symmetry of the form $\lvt$, that is the symmetry leaving $\det(\mcX)$ invariant, %%@
is a real form of $\esi$ as we briefly review here. We closely follow
  references~\cite{Man2,Wang,Man4,Man5,Wang2} within which, in particular, the means %%@
of accommodating, and employing, the non-associative property of the octonions is %%@
described in detail. 

   As a generalisation from the $\sltc \equiv \spot$ Lorentz transformations of %%@
equation~\ref{vinhinc} a set of
   $2 \times 2$ matrix actions with $M \inn \sltwoo \equiv \spotn$  can be embedded in %%@
the upper-left corner of $3 \times 3$ matrices ${\mathcal M}$ to obtain a conjugation %%@
action for the $3 \times 3$ case $R: \mcX \to \mcM \mcX \mcM^{\dag}$ with:
\begin{equation}
   \label{mxm3}
         \mcM \mcX \mcM^{\dag}     = 
		\left( \begin{array}{c|c} 
        \,\,\,\, M                \begin{array}{cc} &  \\  &  \end{array} \!\!\!   &
        \,     0  \begin{array}{cc} &  \\  &  \end{array} \!\!\!\!\!\!\!\!\!\! 
				                         \\  \hline
        \,\,\,\,\,\,  0  \!\! \begin{array}{cc}        &   \end{array}   &	  
		\,  1      \end{array}  \right)  \!\!
		\left( \begin{array}{c|c} 
        \,\,\,\, X                \begin{array}{cc} &  \\  &  \end{array} \!\!\!   &
        \,  \theta  \begin{array}{cc} &  \\  &  \end{array} \!\!\!\!\!\!\!\!\!\! 
				                         \\  \hline
        \,\,\,\,\,\, \theta^{\dagger} \!\! \begin{array}{cc}        &   \end{array}   %%@
&	  
		\,  n      \end{array}  \right)  \!\!
		\left( \begin{array}{c|c} 
        \,\,\,\, M                \begin{array}{cc} &  \\  &  \end{array} \!\!\!   &
        \,  0  \begin{array}{cc} &  \\  &  \end{array} \!\!\!\!\!\!\!\!\!\! 
				                         \\  \hline
        \,\,\,\,\,\, 0 \!\! \begin{array}{cc}        &   \end{array}   &	  
		\,  1      \end{array}  \right)^{\!\!\mbox{\large $\dag$}}  =   
		\left( \begin{array}{c|c} 
       \!  MXM^{\dag}   \!\!\!\!\! 
	           \begin{array}{cc} &  \\  &  \end{array} \!\!\!   &
       \!   M\theta  \!\!\! \begin{array}{cc} &  \\  &  \end{array} %%@
\!\!\!\!\!\!\!\!\!\! 
				                         \\  \hline
        \,\, \theta^{\dagger}M^{\dag} \!\!\!\!\!\! \begin{array}{cc}        &   %%@
\end{array}   &	  
		  \,  n      \end{array}  \right)  
\end{equation}
  This expression contains the vector $X \to R(X) = MXM^{\dag}$, spinor $\theta \to %%@
R(\theta) = M\theta$ and scalar $n \to 1 n$ representations of $\sltwoo$, each %%@
transforming in the appropriate way with the form of the action $R$ determined %%@
correspondingly. These transformations respect the vector and spinor block structure %%@
described in equation~\ref{xoct3}.

  With three natural ways to embed the transformations $\sltwoo \subset \sltho$ the %%@
action depicted in equation~\ref{mxm3} is denoted as `type 1' 
  The form of $\htho$ matrices transforming under the type 1 $\sltc$ and $\sltwoo$ %%@
actions is compatible with the isomorphism of vector spaces (\cite{Baez1} p.30):
\begin{eqnarray}
    \htho  &   \cong    &   \rrr \oplus  \htwo  \oplus  \ooo^2  \label{horhoo} \\ 
	  \nonumber \\    
	\left( \begin{array}{cc} \!\!\!\!
       \left( \,\,\,\,\,\,\, X \begin{array}{cc} &  \\  &  \end{array} \!\!\!  \right) %%@
&
    \!\!\!\!\! \left(  \theta \begin{array}{cc} &  \\  &  \end{array} \!\!\!\!\! %%@
\!\!\!\!\! \right) 
				                         \\ 
	\!\!\!	\left( \,\,\,\,\,\,\, \theta^{\dagger} \begin{array}{cc} &   \end{array} %%@
\!\!\!  \right)  &			  
					    \!\!\!\!\! n \end{array} \!\!\!\! \right)     
	      &  \to  &   (n , \quad \; X, \quad \; \theta)  \label{xinmcx}    \\   %%@
\nonumber   \\
	  \mathbf{27}_{\mathrm{E}_6}  &  \to  &  (\mathbf{1} + \mathbf{10} + %%@
\mathbf{16})_{\mathrm{Spin}^+(1,9)}  \label{esidecom}
\end{eqnarray}
  The three parts of this decomposition are respectively the scalar, vector and spinor %%@
representations of the 10-dimensional spacetime symmetry group $\sootn$, for which the %%@
covering group is $\spotn \equiv \sltwoo$.
  The object $\theta = \binom{c}{\bar{b}}$, from equation~\ref{xoct3}, corresponds to %%@
the Majorana-Weyl spinor representation, also denoted as $\mathbf{16}$.

  With the group $\sltwoo$ being 45-dimensional and with three distinct types of %%@
embedding this implies to a total of $45 \times 3 = 135$ determinant preserving %%@
actions on the space $\htho$. The collective symmetry of these actions is described  %%@
in terms of vector fields on the tangent space to $\htho$:
  \begin{equation}
\label{rdot}
   \dot{R} = \frac{\partial \: ( R(\alpha) \mcX  )}
      {\partial \alpha} \Big|_{\alpha = 0} \; \inn \; T\htho
\end{equation}
 where $\alpha \inn \rrr$ parametrises a given action. By studying the linear %%@
dependences of the resulting 135 vector fields it can be %%@
shown~\cite{Wang,Man4,Man5,Wang2} how
 a linearly independent basis of 78 actions emerges which fully describes $\esi \equiv %%@
\sltho$ as the  determinant preserving  transformations on $\htho$. The entire group %%@
is then described in terms of the actions of  matrices $\mcM$ on the space $\htho$, %%@
with the preferred basis for the Lie algebra represented on $T\htho$ reproduced below %%@
in table~\ref{prefbas}.
\begin{table}[htbp]
\centering
\begin{tabular}{|ccc|r|}
 \hline  
 \multicolumn{3}{|l|}{Category 1: Boosts}    & \#  \\ \cline{4-4}
   $\qquad  \dot{B}_{t \pqg z}^{1} \qquad$
	 & $\qquad \dot{B}_{t \pqg x}^{1} \qquad$
	   &  $\qquad \dot{B}_{t \pqg q}^{1} \qquad$ &  $9$  \\
	$\dot{B}_{t \pqg z}^{2}$ & $\dot{B}_{t \pqg x}^{2}$ & $\dot{B}_{t \pqg q}^{2}$ &  %%@
$9$  \\
                               & $\dot{B}_{t \pqg x}^{3}$ & $\dot{B}_{t \pqg q}^{3}$ &  %%@
$8$  \\
 \hline
 \multicolumn{3}{|l|}{Category 2: Rotations} &   \\
    $\dot{R}_{x \pqg q}^{1}$ & $\dot{R}_{x \pqg z}^{1}$ & $\dot{R}_{z \pqg q}^{1}$ & %%@
$15$  \\
	                           & $\dot{R}_{x \pqg z}^{2}$ & $\dot{R}_{z \pqg q}^{2}$ &  %%@
$8$  \\
	                           & $\dot{R}_{x \pqg z}^{3}$ & $\dot{R}_{z \pqg q}^{3}$ &  %%@
$8$  \\
 \hline
 \multicolumn{3}{|l|}{Category 3: Transverse Rotations} &  \\ 
    $\dot{A}_q$                & $\dot{G}_q$                & $\dot{S}_{q}^{1}$        %%@
& $21$  \\
 \hline
 \multicolumn{3}{|r|}{Total Generators}    &    78  \\
 \hline
  \end{tabular}
  \caption{\setb The complete basis for the Lie algebra of $\esi$, in terms of tangent %%@
vector fields on $T\htho$, reproduced from (\protect\cite{Wang} p.177, table A.1).
 The superscripts denote the `type'. The subscript $q$ denotes any of the seven %%@
imaginary octonion units $\{i,j,k,\kl,\jl,\il,l \}$.}
\label{prefbas}
\end{table}

  Here all 78 generators are explicitly determined and listed in tables~\ref{lbrota} %%@
and \ref{ltrota} in the appendix of this paper, for the category $\{1,2\}$ and 3 %%@
transformations respectively, as tangent vector fields $\dot{R} \inn T\htho$ which, %%@
from equation~\ref{xoct3}, are of the form:
\begin{equation}
 \label{ththo}
  \dot{R} = 
   \left( \begin{array}{ccc}
       \dot{p} & \dot{\bar{a}} & \dot{c}  \\
       \dot{a} &   \dot{m}     & \dot{\bar{b}}        \\
 \dot{\bar{c}} &   \dot{b}     & \dot{n} 
          \end{array}  \right)   \inn T\htho 
\end{equation}

 The Lie algebra commutator, which determines the structure constants of the $\esi$ %%@
Lie algebra, for any two elements $\dot{R}_1, \dot{R}_2$ is defined through the action %%@
of the respective one-parameter subgroups $R_1(\alpha)$ and $R_2(\alpha)$ at any point %%@
$\mcX \inn \htho$:
\begin{equation}
 [\dot{R}_2,\dot{R}_1] \; = \; \frac{\partial}{\partial (\alpha^2)}
    [R_2(-\alpha)\,\scirc\, R_1(-\alpha)\,\scirc\,
	   R_2(\alpha)\,\scirc\, R_1(\alpha)\,\mcX] \Big|_{\alpha = 0}
\label{rrbrac}
\end{equation}

  The full $\esi$ Lie algebra commutation table is available in \cite{Wang}, for which %%@
the full set of $(78 \times 78 - 78)/2 = 3003$ independent entries were found by %%@
computer program. For this paper sign and other conventions have been tuned both for %%@
internal consistency and for consistency with the entries of the full $\esi$ algebra %%@
table $\cite{Wang}$. (In this paper  `$\esi$' may refer to either the Lie group or the %%@
Lie algebra, depending on the context, with a similar convention for the other %%@
exceptional Lie groups).%%@

\section{$\esi$ Symmetry Breaking Structure}
\label{chapesb}

 We can identify the Lorentz 4-vector $\bv_4=(v^0,v^1,v^2,v^3)\equiv \bh_2$ in the %%@
upper left-hand $2\times 2$ matrix embedded within the larger $3\times 3$ matrices in %%@
$\htho$, as was the case for $\htwc \subset \hthc$ in equation~\ref{vinhinc}. 
 However here the preferred subspace $\ccc\subset \ooo$ basis is taken to be $\{1,l\}$ %%@
for $\bv_4$, with $l$ one of the imaginary octonion units introduced in %%@
equation~\ref{octa}, as indicated in equation~\ref{hinhtho}.
The relation $\mbox{det}(\mcX) = 1$ with $\mcX \inn \htho$ is preserved under %%@
operations of $\sltc$ representing the Lorentz group upon this space as: 
\vspace{5pt}
\begin{equation}
   \label{hvvv}
         \mcX \; \to \;
		\left( \begin{array}{c|c} 
        \,\,\,\,\, S   \!      \begin{array}{cc} &  \\  &  \end{array} \!\!\!   &
        \,     0  \begin{array}{cc} &  \\  &  \end{array} \!\!\!\!\!\!\!\!\!\! 
				                         \\  \hline
        \,\,\,\,\,\,  0  \!\! \begin{array}{cc}        &   \end{array}   &	  
		\,  1      \end{array}  \right)  
		\left(\! \begin{array}{cc|c} 		
            h^{00}    & h^{01}\! + \overline{a}(6)   &   c  \\
		  h^{10}\! + a(6)   &    h^{11}         & \overline{b}  \\         	
               \hline										 
         \overline{c}     &   b  &  n      \end{array}  \right)  
		\left( \begin{array}{c|c} 
        \,\,\,\,\, S^{\dag}  \!    \begin{array}{cc} &  \\  &  \end{array} \!\!\!   &
        \,  0  \begin{array}{cc} &  \\  &  \end{array} \!\!\!\!\!\!\!\!\!\! 
				                         \\  \hline
        \,\,\,\,\,\, 0 \!\! \begin{array}{cc}        &   \end{array}   &	  
		\,  1      \end{array}  \right) 
\end{equation}
\vspace{5pt}
\begin{equation}
 \label{hinhtho}
  \mbox{with}    
		 \qquad \qquad
         \begin{array}{cc} h^{00}=v^0+v^3,\quad  &  h^{01}=v^1-v^2l  \\
		                   h^{10}=v^1+v^2l,\quad &  h^{11}=v^0-v^3   \end{array}
					\qquad \qquad \qquad	   
 \end{equation}
   with $S\inn\sltc$, and with `$1$' describing the identity transformation in the %%@
trivial 1-dimensional representation of this group, acting upon the components of %%@
$\mcX$ of equation~\ref{xoct3}. This action preserves the value of det$(\bh_2) = h^2$, %%@
as it is simply the transformation of equation~\ref{hshs}, as well as leaving %%@
$\mbox{det}(\mcX) = 1$ invariant.  In equation~\ref{hvvv} $a(6)$ denotes the %%@
6-dimensional imaginary part of $a\inn \htho$ of equation~\ref{xoct3}, that is %%@
excluding the real $a_1 = v^1$ and imaginary $a_8l = v^2l$ components of $a\inn \ooo$ %%@
which are associated with the external 4-vector $\bv_4 \inn \TM_4$.
Since the $\sltc$ actions, based on the $\{1,l\}$ complex subspace, are embedded in %%@
the `type 1' location  this group will be denoted $\sltc^1$. This describes the
 Lorentz subgroup for 
4-dimensional spacetime  generated by the
 subset of
 Lie algebra elements in $\esi$ from table~\ref{prefbas}:
\begin{equation}
   \{\dot{B}_{t \pqg z}^{1}, \dot{R}_{x \pqg l}^{1}, \dot{B}_{t \pqg x}^{1},
     \dot{B}_{t \pqg l}^{1}, \dot{R}_{x \pqg z}^{1}, \dot{R}_{z \pqg l}^{1} \}
	 \label{extlor6}
\end{equation}

 With the group action of $\sltc^1$ on $\htho$ in equation~\ref{hvvv}  embedded within %%@
the type 1 group action of $\sltwoo^1$ on the same space as displayed in %%@
equation~\ref{mxm3} we can write:
  \begin{equation}
   \label{sltcinesi}
    \soot \equiv \sltc^1 \subset \sltwoo^1 \subset \sltho \equiv \esi
  \end{equation}
   where the first `$\equiv$' strictly applies at the Lie algebra level.
 This shows explicitly how the action of the Lorentz group may be embedded within the %%@
higher symmetry group $\esi$ acting on the space $\htho$. The direct physical %%@
interpretation of the former symmetry as an action on the external spacetime manifold %%@
$M_4$ (as originally depicted in figure~\ref{mtogmaphr}(b))  provides a direct source %%@
for the breakdown of the latter symmetry.

The two-sided $\sltc^1$ action on $\htwo$  in equation~\ref{hvvv} only transforms the %%@
real diagonal entries $h^{00}$ and $h^{11}$ together with the $h^{10} = a_1 + a_8l$ %%@
and $h^{01} = a_1 - a_8l$ components of $a\inn \ooo$. The six components of $a(6) \inn %%@
\mbox{Im}(a)$ remain invariant as may be deduced from the form of the six $\sltc^1$ %%@
generators in table~\ref{lbrota}.
 (This is equivalent to the invariance of $\ul{\bv}_6$ under $\soot$ for the model of %%@
figure~\ref{mtogmaphr}).
 Of the additional 17 components in $\htho$ the real diagonal $n$ entry is also %%@
invariant, as is clear from equation~\ref{hvvv}, while \textit{all} 16 components of %%@
$b,c \inn \ooo$ transform non-trivially under the one-sided $\sltc^1$ action.

The spinor $\theta_l = \binom{c}{\bar{b}}_{\! l} \inn \ccc^2$ will denote the %%@
$\{1,l\}$ components of $c$ and $\bar{b}$ in $\theta$, that is $\binom{c}{\bar{b}}\inn %%@
\ooo^2$ restricted to the $\{1,l\}$ complex subspace. By comparison with %%@
equation~\ref{vinhinc} this object transforms as a left-handed Weyl spinor  $\psi_L = %%@
\theta_l$ under the $\sltc^1$ action in equation~\ref{hvvv}. Here we take this
 $S \inn \sltc^1$ action on $\theta_l$ to \textit{define} the left-handed
  spinor representation.

  Considering the three quaternionic subspaces with the base units $\{1, l, i, \il\}$, %%@
$\{1, l, j, \jl\}$ and $\{1, l, k, \kl\}$ it can be seen through explicit calculation  %%@
that the original full octonionic spinor $\theta = \binom{c}{\bar{b}}$, with 16 real %%@
components, reduces to a total of four left-handed Weyl spinors under the action of %%@
$\sltc^1$, namely:
\begin{equation}
   \theta_{l} \! = \! 
        \left( \begin{array}{c} c_1 + c_8l  \\
		                        b_1 - b_8l      \end{array}  \right) \!\! , \;\,
    \theta_i \! = \! \left(\!\! \begin{array}{c}  c_7\il + c_2i \\
		                     - b_7\il - b_2i     \end{array} \!\! \right) \!\! , \;\,
	 \theta_j \! = \! \left(\!\! \begin{array}{c}  c_6\jl + c_3j \\
		                     - b_6\jl - b_3j     \end{array} \!\! \right) \!\! , \;\,
	 \theta_k \! = \! \left(\!\! \begin{array}{c}  c_5\kl + c_4k \\
		                     - b_5\kl - b_4k     \end{array} \!\! \right)
		\label{thcth234} 
\end{equation}
 with the respective components of each  transforming in an identical manner.
 Overall the decomposition of the $\mathbf{27}$ representation of $\esi$ under the %%@
subgroup $\spotn$ of equation~\ref{esidecom} further reduces under the subgroup %%@
$\sltc^1$ as summarised in table~\ref{sltcreps}.

\begin{table}[htbp]
\centering
\begin{tabular}{|c|c|c|}
 \hline
     $\spotn$       &   $\sltc^1$                &    Components     \\  
 \hline		  
    \: $\mathbf{1}$ \, scalar
	    &  $\qquad$  (0,0) $\;$ scalar            &    $n$            \\
	 $\mathbf{10}$ \, vector
	   &  $ \left\{ \begin{array}{rl}  (\fhs,\fhs) &  \mbox{vector}  \\
                   6\times (0,0) & \mbox{scalars}  \end{array}  \right. $  &         
	$	\begin{array}{c}     \bv_4   \\  a(6)  \end{array} $  \\
	 $\mathbf{16}$ \, spinor
	   & $\quad \! 4\times(\fhs,0)\;\,$ spinors   &    $\theta_{l,i,j,k}$   \\ 
   \hline
  \end{tabular}
  \caption{\setb The further decomposition of the $(\mathbf{1} + \mathbf{10} + %%@
\mathbf{16})$ representation of $\spotn \subset \esi$ of equation~\ref{esidecom} under %%@
the subgroup of external 4-dimensional spacetime symmetry $\sltc^1 \subset \spotn$ %%@
actions of equation~\ref{hvvv}, and the corresponding components of $\htho$ %%@
transformed.}
\label{sltcreps}
\end{table}

  As a preliminary definition, and in contrast to the external symmetry, the internal %%@
symmetry will be obtained from the set of all $\esi$ actions on $\htho$ which leave %%@
the four components for any $\bv_4 = (v^0,v^1,v^2,v^3)$ in equation~\ref{hinhtho} %%@
(that is, $\bh_2 \inn \htwc$ of equation~\ref{hvvv}) invariant. These components, %%@
including $v^2 \equiv a_8$ associated with the imaginary unit $l$ of $a\inn \ooo$, can %%@
also be expressed in the combination $(p,m,a_1,a_8)$ with respect to the %%@
parametrisation of equations~\ref{xoct3} and \ref{octa}.  The corresponding symmetry %%@
group is  complementary to the actions of $\sltc^1$ and will be denoted $\stab$ as the %%@
stability group of all vectors $\bv_4 \inn \TM_4$. By inspection from %%@
tables~\ref{lbrota} and \ref{ltrota}, for the 78 elements in the preferred basis for %%@
the Lie algebra of $\esi$ defined on the space $T\htho$, the group $\stab$ is %%@
generated by the 31 elements listed in table~\ref{stabset}.

\begin{table}[htbp]
\centering
\begin{tabular}{|l|r|}
 \hline  
      Category 1 and 2: Boosts and Rotations   & \#  \\ \cline{2-2}
	  $\qquad (\dot{R}_{x \pqg z}^{2} - \dot{B}_{t \pqg x}^{2}),	
   \qquad   (\dot{R}_{x \pqg z}^{3} + \dot{B}_{t \pqg x}^{3})$	  &   2  \\
      $\qquad (\dot{R}_{z \pqg q}^{2} + \dot{B}_{t \pqg q}^{2}),
   \qquad   (\dot{R}_{z \pqg q}^{3} - \dot{B}_{t \pqg q}^{3})$    &  14  \\
 \hline
     Category 3: Transverse Rotations  &  \\ 
    $\qquad \dot{A}_q$,  \quad   $\dot{G}_l$,  \quad $\dot{S}_{l}^{1}$      &   9  \\
	$\qquad (\dot{G}_q + 2\dot{S}_{q}^{1}) \;\;\; q= \{i,j,k,\kl,\jl,\il \}$   &   6  %%@
\\
 \hline
 \multicolumn{1}{|r|}{Total}    &    31  \\
 \hline
  \end{tabular}
  \caption{\setb The Lie algebra generators of the 31 dimensional group $\stab$. The %%@
subscript $q$ denotes any of the seven imaginary octonion units $\{i,j,k,\kl,\jl,\il,l %%@
\}$ unless stated otherwise.}
\label{stabset}
\end{table}

Of particular interest is the $\suth$ subgroup associated with the 8 generators  %%@
$\{\dot{A}_q, \dot{G}_l\}$ 
  (as also described in \cite{Wang} pp.115 and 136, following \cite{Gunay}
  in relation to the $\suth$ colour symmetry).
 This $\suth$ is defined in terms of the transverse rotations acting on the octonion %%@
space $\ooo$ alone as the subgroup $\suth \subset \gt$ of the octonion automorphism %%@
group that leaves one imaginary unit, here $l$, invariant.
The corresponding Lie algebra $\sutha$, described by the set of 8 generators %%@
$\{\dot{A}_q, \dot{G}_l\}$, as transformations of $\esi$ on the full space $\htho$, %%@
act on each of the octonion elements $a,b,c \inn \ooo$ in the same way (as described %%@
in the table~\ref{ltrota} caption) leaving invariant the complex $\{1,l\}$ subspaces, %%@
and as elements  of table~\ref{stabset}
   identified within $\staba$ may be provisionally associated with the colour %%@
su(3)$_c$ of the Standard Model for the present theory.
 This algebra is also independent of $\sltc^1$ in terms of the Lie bracket %%@
composition, that is $\lbrack X,Y \rbrack = 0$ for all $X\inn \sltca^1$ and $Y \inn %%@
\sutha_c$, and hence we have the semi-simple subgroup:
\begin{equation}
  \label{esisubss}
    \sltc^1 \times \suth_c \subset \esi
\end{equation}

 The $\{\dot{A}_q, \dot{G}_l\}$ algebra elements, as transformations on the space %%@
$\htho$ mix the components of the $\spotn$ spinor $\theta = \binom{c}{\bar{b}} \inn %%@
\ooo^2$. For example the tangent vector field $\dot{A}_i$ on the $\binom{c}{\bar{b}}$ %%@
components of $\htho$, obtained from table~\ref{ltrota}, are:
\begin{eqnarray} 
 \!\!\!\!\!\!\!\!\!  
    \dot{A}_i:\left( \begin{array}{c} \dot{c} \\ \dot{\bar{b}} \end{array} \right)
	 \!  & \! = \! & \!
	 \left(  \begin{array}{cccc} 
 \dot{c}_1+\dot{c}_8l,\sb &+\dot{c}_7\il+\dot{c}_2i,\sb &+\dot{c}_6\jl+\dot{c}_3j,\sb 
   &+\dot{c}_5\kl+\dot{c}_4k \\
 \dot{b}_1-\dot{b}_8l,\sb &-\dot{b}_7\il-\dot{b}_2i,\sb &-\dot{b}_6\jl-\dot{b}_3j,\sb  
  &-\dot{b}_5\kl-\dot{b}_4k
  	   \end{array} \right)  \nonumber \\
	 \! & \! = \! & \!
	 \left(  \begin{array}{cccc} 
  0 + 0l, \sb\;\:\, & + 0 \il+0 i, \sb\;\;\, &-{c_5}\jl{-c_4}j, \sb\:\,  %%@
&+{c_6}\kl+{c_3}k \\
  0 - 0l, \sb\;\:\, & - 0 \il-0 i, \sb\;\;\, &+{b_5}\jl{+b_4}j, \sb\:\,  %%@
&-{b_6}\kl-{b_3}k
  	   \end{array} \right)
  \label{aioncb}
\end{eqnarray}

   The components here have been ordered to match those of the four left-handed Weyl %%@
spinors $(\theta_l, \, \theta_i, \, \theta_j, \theta_k)$ of equation~\ref{thcth234}. %%@
The fact that each real component of $c$ transforms in the same way as the %%@
corresponding component of $b$ is expected since $\suth_c$ acts on each of $a,b,c \inn %%@
\ooo$ in precisely the same way. However, it is also noted that the action $\dot{A}_i$ %%@
in equation~\ref{aioncb}  respects the 4-way spinor decomposition.

 The extraction of the components of a spinor $\theta$ into a matrix of real numbers %%@
will be denoted by $\lbrack \theta \rbrack$. For example, from equation~\ref{thcth234} %%@
the spinor $\theta_i$ can be mapped to the $2\times 2$ matrix of real numbers $\lbrack %%@
\theta_i \rbrack = \binom{ c_7 \;\;\; c_2}{-b_7 \, -b_2}$ (with components ordered to %%@
match those of the spinor $\theta_l$ under $\sltc^1$ transformations, as described for %%@
equation~\ref{thcth234}). Using this notation, and with $0_2$ representing the %%@
$2\times 2$ zero matrix, the tangent vectors of all eight generators $\{\dot{A}_q, %%@
\dot{G}_l\}$ of SU(3)$_c$, including $\dot{A}_i$ from equation~\ref{aioncb}, on the %%@
spinor space $\theta \inn \ooo^2$ are listed in table~\ref{suttan} alongside the %%@
actions of the Gell-Mann matrices $\lambda_{\alpha}$, using the correspondence of %%@
(\protect\cite{Wang} p.137, table 4.5), on the vectors $\bu\inn\ccc^3$. On the %%@
left-hand side            
the elements $\{\dot{A}_q, \dot{G}_l\}$ are already expressed as tangent vectors, %%@
while on the right-hand side the tangents are  
obtained by matrix multiplication of the $\lambda_{\alpha}$ into $\bu \inn \ccc^3$.

\begin{table}[htbp]
\centering
 \hspace*{-12pt}
\begin{tabular}{|c|}
\hline
 $ \begin{array}{c@{(}rrrr@{)}ccc@{(}ccc@{)}}
   &   \lbrack \dot{\theta}_{l} \rbrack, & \lbrack \dot{\theta}_i \rbrack, & 
	   \lbrack \dot{\theta}_j \rbrack,      & \lbrack \dot{\theta}_k \rbrack 
	 & &  & &   \dot{u}_1, & \dot{u}_3, & \dot{u}_3 
\vspace{10pt} \\ 
	 \dot{A}_i = \quad &   0_2, & 0_2, &  
	                  -\lbrack {\theta}_k \rbrack, & \lbrack {\theta}_j \rbrack  
	& \qquad \quad \sim \qquad & \lambda_4 \Rightarrow &   &  u_3, &  0, & u_1 \\
     \dot{A}_{\il} = \quad &   0_2, & 0_2, &  
	                  \lbrack {l\theta}_k \rbrack, & \lbrack {l\theta}_j \rbrack  
	& \qquad \quad \sim \qquad & \lambda_5 \Rightarrow &   &   -iu_3, &  0, & iu_1  
 \vspace{10pt} \\ 
	 \dot{A}_j = \quad &   0_2, &  \lbrack {\theta}_k \rbrack, &
	                       0_2, &   -\lbrack {\theta}_i \rbrack   
	& \qquad \quad \sim \qquad & \lambda_7 \Rightarrow &   &   0, & -iu_3, & iu_2  \\
	\dot{A}_{\jl} = \quad &   0_2, &  -\lbrack {l\theta}_k \rbrack, &
	                       0_2, &   -\lbrack {l\theta}_i \rbrack   
	& \qquad \quad \sim \qquad & \lambda_6 \Rightarrow &   &   0, & u_3, & u_2  
 \vspace{10pt} \\ 
	 \dot{A}_k = \quad &   0_2, &  -\lbrack {\theta}_j \rbrack, &
	                        \lbrack {\theta}_i \rbrack, & 0_2   
	& \qquad \quad \sim \qquad & \lambda_1 \Rightarrow &   &   u_2, & u_1, & 0 \\
	\dot{A}_{\kl} = \quad &   0_2, &  \lbrack {l\theta}_j \rbrack, &
	                        \lbrack {l\theta}_i \rbrack, & 0_2   
	& \qquad \quad \sim \qquad & \lambda_2 \Rightarrow &   &   -iu_2, & iu_1, & 0 
 \vspace{10pt} \\ 
	 \dot{A}_l = \quad &   0_2, &  \lbrack {l\theta}_i \rbrack, &
	                        -\lbrack {l\theta}_j \rbrack, & 0_2   
	& \qquad \quad \sim \qquad & \lambda_3 \Rightarrow &   &   u_1, & -u_2, & 0 \\
	 \dot{G}_l = \quad &   0_2, &  \lbrack {l\theta}_i \rbrack, &
	                    \lbrack {l\theta}_j \rbrack, & -2\lbrack {l\theta}_k \rbrack  
	& \qquad \quad \sim \qquad & \lambda_8 \Rightarrow & \!\! \frac{1}{\sqrt{3}}  &   %%@
u_1, & u_2, & \!\! -2u_3 \\
 \end{array}
 $	 
  \\
 \hline
  \end{tabular}
  \caption{\setb The tangent vector generators for the SU(3)$_c$ representations on %%@
$\ooo^2$ and $\ccc^3$.  The column vectors of  $\ccc^3$ are displayed as a row vectors %%@
for convenience in the table.}
\label{suttan}
\end{table}

   In table~\ref{suttan} a term such as $\lbrack l\theta_i \rbrack$ denotes %%@
multiplying the spinor $\theta_i$ on the left by $l$ before extracting the %%@
coefficients of $l\theta_i$ with the $\mbox{Im}(\ooo)$ units ordered as in %%@
equation~\ref{thcth234}. This notation is used to isolate the mixing effect on the %%@
real number coefficients, with care for the joint effects of the division algebra %%@
composition as well as matrix algebra composition. 
 The internal SU(3)$_c$ symmetry action on the left-hand side of table~\ref{suttan} %%@
dovetails neatly with the external $\sltc^1$ spinor structure of %%@
equation~\ref{thcth234}.
 The mixing action of SU(3)$_c$ in table~\ref{suttan} takes a form summarised as:
\begin{equation}
    \theta  \;  = \; (\, \theta_l, \quad
	  \underbrace{ \theta_i, \quad \theta_j, \quad \theta_k}_{\mbox{\small SU(3)$_c$ %%@
action}} \, )
	  \label{suthth}
\end{equation}
   which implies that as a gauge theory the $\suth_c$ internal symmetry will mediate %%@
interactions between the
    Weyl spinors $\theta_i, \theta_j, \theta_k$, transforming under the fundamental %%@
representation, which in turn will hence be identified with the three colour degrees %%@
of freedom of the quark states. On the other hand the invariance of $\theta_l$,
 transforming under the trivial representation of SU(3)$_c$, suggests that these %%@
components should be associated with the leptonic sector of the Standard Model (with %%@
the subscript $l$ originating from the $\{1,l\}$ base units for $\theta_l$ also then %%@
serving as a mnemonic for its leptonic character).

Of the 31 generators for the internal symmetry group $\stab$ listed in %%@
table~\ref{stabset} there is a $(31-8)=23$-dimensional set which as a vector space is %%@
independent of the internal SU(3)$_c$ generators. However, of these 23 elements %%@
$\dot{S}_{l}^{1}$  as the \textit{only} $\esi$ Lie algebra generator of $\stab$ which %%@
is independent of both $\sltc^1$ and $\suth_c$.
 Hence of the many possible $\uo \subset \esi$ subgroups the internal $\uo$ generated %%@
by $\dot{S}_{l}^{1}$ is a natural candidate to consider for the $\uo_Q$ component of %%@
the Standard Model gauge symmetry group associated with electromagnetism.
 From table~\ref{ltrota} it can be seen that the generator $\dot{S}_{l}^{1}$ impacts %%@
on all 8 real components of both $c$ and $\bar{b}$ of $\theta \inn \ooo^2$. In fact, %%@
and in comparison with equation~\ref{aioncb}, the tangent vector  $\dot{S}_{l}^{1}$ on %%@
the spinor components $\theta = \binom{c}{\bar{b}}$ is given explicitly by:
\begin{eqnarray} 
   \!\!\!   \dot{S}_l^1 \! : \!\left( \! \begin{array}{c} \dot{c} \\ \dot{\bar{b}} %%@
\end{array} \! \right) \!\! & \!\!\! = \!\!\! & \!\!
	 \left( \;\!  \begin{array}{cccc} 
 \dot{c}_1+\dot{c}_8l, \quad\;\, &+\dot{c}_7\il+\dot{c}_2i,  \quad\;\, 
                 &+\dot{c}_6\jl+\dot{c}_3j, \quad\;\,  &+\dot{c}_5\kl+\dot{c}_4k \\
  \dot{b}_1-\dot{b}_8l, \quad\;\, &-\dot{b}_7\il-\dot{b}_2i, \quad\;\,
                 &-\dot{b}_6\jl-\dot{b}_3j, \quad\;\, &-\dot{b}_5\kl-\dot{b}_4k
  	   \end{array} \;\! \right)  \nonumber \\
	\!\!  & \!\!\! = \!\!\! & \!\!
	 \left(  \!\! \begin{array}{cccc} 
  -\frac{3}{2}c_8 + \frac{3}{2}c_1l, & +\fh c_2\il- \fh c_7 i, & +\fh c_3 \jl - \fh %%@
c_6 j, & +\fh c_4 \kl - \fh c_5 k \\
  \frac{3}{2}b_8 + \frac{3}{2}b_1l, & -\fh b_2\il+ \fh b_7 i, & -\fh b_3 \jl + \fh b_6 %%@
j,  & -\fh b_4 \kl + \fh b_5 k
  	   \end{array} \!\! \right) \quad\;\;   \label{sloncb}    \\ & & \nonumber \\
 \!\!\! \mbox{with}\quad \!\! \lbrack \dot{\theta} \rbrack 
    \;\;\!\! & \!\!\! = \!\!\! & \;\!\!
  \left(\quad +\frac{3}{2}\;\lbrack l\theta_l \rbrack,
        \qquad\quad -\frac{1}{2}\;\lbrack l\theta_i \rbrack,    
        \qquad\quad -\frac{1}{2}\;\lbrack l\theta_j \rbrack,
		\qquad\quad -\frac{1}{2}\;\lbrack l\theta_k \rbrack \quad \right)  \qquad   
  \label{slonthet}
\end{eqnarray}
  which may be compared with the su(3)$_c$ action on $\theta$ in equation~\ref{aioncb} %%@
and table~\ref{suttan}.
   Here the two components of $c\inn \ooo$ \textit{within} each of the four Weyl %%@
spinors are mixed, and similarly for the corresponding pair of $\bar{b}\inn \ooo$ %%@
components, with no mixing of components \textit{between} different spinors. This is %%@
consistent with the nature of the electromagnetic interaction which does not transform %%@
between different fermion types.

 A further observation from equation~\ref{slonthet} regards the factor of %%@
$\frac{3}{2}$ found for the $\theta_l$ spinor in contrast to the factors of $\fh$ %%@
aligned with the three remaining spinors $\theta_i, \theta_j, \theta_k$. Hence, with %%@
$\dot{S}_{l}^{1}$ provisionally associated with electromagnetism and by comparison %%@
with equation~\ref{suthth}, the apparent  `electromagnetic charge' assigned to the %%@
leptonic sector is \textit{three times} larger than that assigned to the quark sector. %%@
Associating $\theta_i, \theta_j, \theta_k$ with the three colour states of a $d$-quark %%@
this observation in principle accounts for the `fractional charge' of magnitude %%@
$\frac{1}{3}$ as theoretically ascribed and empirically confirmed for $d$-quark states %%@
relative to the electron charge. Based on this observation we introduce the notation:
\begin{equation}
  \label{sbardot}
      \Sbard^a_l = \frac{2}{3}\dot{S}^a_l
\end{equation}
    (for $a=1,2,3$) such that the above charge values $\frac{3}{2}$ and $\frac{1}{2}$ %%@
are normalised to $1$ and $\frac{1}{3}$ under $\Sbard^1_l$, representing the generator %%@
of $\uo_Q$,
for ease of comparison with the Standard Model convention for which the electron %%@
charge is $-1$. The `bar' through $\Sbard^1_l$ is a mnemonic symbol for this %%@
normalisation of fractional charges relative to the $e^-$ charge.

Hence the subgroup in equation~\ref{esisubss} may be augmented to: 
\begin{equation}
  \label{esisubssu}
    \sltc^1 \times \suth_c \times \uo_Q \subset \esi
\end{equation}
  with the internal group $\suth_c \times \uo_Q$ generated by $\{\dot{A}_q, \dot{G}_l, %%@
\Sbard^1_l \} \inn \staba$. The action of this larger internal symmetry on the four %%@
$\sltc^1$ spinors also augments equation~\ref{suthth} as:
\begin{eqnarray}
    \theta  &  = & (\, \theta_l, \quad
	  \underbrace{ \theta_i, \quad \theta_j, \quad \theta_k} \, )  \nonumber \\
  \suth_c & : &  \;\;  \mathbf{1} \qquad \qquad  \mathbf{3} \label{suthuoth} \\
    \uo_Q   & : &   +1 \;\; -\!{\mbox{\small{$\frac{1}{3}$}}}
	     \;\; -\!{\mbox{\small{$\frac{1}{3}$}}}  \;\; %%@
-\!{\mbox{\small{$\frac{1}{3}$}}}
	  \nonumber
\end{eqnarray}

  With the generator $\Sbard^1_l$ hence associated with electromagnetic charge it is %%@
instructive to consider this action on the full set of $\htho$ components. From %%@
table~\ref{ltrota} the diagonal components of $\Sbard^1_l$ are trivial, with $\dot{p} %%@
= \dot{m} = \dot{n} = 0$ in equation~\ref{ththo} for this generator, while the action %%@
on the remaining components $a,b,c \inn \ooo$, via equation~\ref{sbardot}, may be %%@
summarised as:
\begin{equation}
  \label{sbcharge}
   \Sbard_l^1 \; = \;
   \left( \begin{array}{c} \dot{a} \\ \dot{b} \\ \dot{c} \end{array}
              \right) \; = \; 
			  \left( \!\!\! \begin{array}{rcr}
			   0\,l\,a_{1,l} &\!\! + \!\!& \frac{2}{3}\,l\,a(6)  \\
              -1\,l\,b_{1,l} &\!\! - \!\!& \frac{1}{3}\,l\,b(6)  \\
	          +1\,l\,c_{1,l} &\!\! - \!\!& \frac{1}{3}\,l\,c(6)
			  \end{array} \! \right) 
\end{equation}
	where $a_{1,l} \equiv (a_1 + a_8l)$ and $a(6) \equiv 
	(a_7\il+a_2i+a_6\jl+a_3j+a_5\kl+a_4k)$, with similar expressions for $b$ and $c$, %%@
following the component order of the spinors in equation~\ref{thcth234}.
 The same definition of $a(6)$ is implied in equation~\ref{hvvv}.
 By comparison with the above discussion leading to equation~\ref{suthuoth} the %%@
expression for $\Sbard_l^1$ in equation~\ref{sbcharge} incorporates `charges' of 0 and %%@
$\frac{2}{3}$ for the $\dot{a}$ components, that is we have:
\begin{eqnarray}
    a  &  = & (\, a_{1,l}, \;
	  \underbrace{ a_{\il,i}, \; a_{\jl,j}, \; a_{\kl,k} } \, )  \nonumber \\
  \suth_c & : &  \;\;\;  \mathbf{1} \qquad \qquad \:\! \mathbf{3} \label{suthuoa} \\
    \uo_Q   & : &  \;\;\; 0 \;\;\, +\!{\mbox{\small{$\frac{2}{3}$}}}
	     \;\;\, +\!{\mbox{\small{$\frac{2}{3}$}}}  \;\;\, %%@
+\!{\mbox{\small{$\frac{2}{3}$}}}
	  \nonumber
\end{eqnarray}
  where the $\suth_c$ action on $a\inn \ooo$ is identical to that on the octonion %%@
components of $\theta = \binom{c}{\bar{b}}$ in equation~\ref{suthuoth}.
 While physical lepton states are invariant under SU(3)$_c$ and are hence  associated %%@
with the Weyl spinor $\theta_l$ in equation~\ref{suthuoth}, the neutrino states are %%@
also invariant under the $\uo_Q$ of electromagnetism, that is with zero charge, and %%@
are provisionally associated with the $a_{1,l}$ components in equations~\ref{sbcharge} %%@
and \ref{suthuoa};
 while a set of $u$-quarks with $\frac{2}{3}$ fractional charges is similarly %%@
associated with the $a(6)$ components.

  However, unlike $\theta=\binom{c}{\bar{b}}$ the $a\inn \htho$ component does %%@
\textit{not} correspond to a set of $\sltc^1$ Weyl spinors, as can be seen from %%@
table~\ref{sltcreps}.
 Further, the `neutrino' components $a_{1,l} = a_1 + a_8l = v^1 + v^2l$ have already %%@
apparently been accounted for as part of the external vector $\bv_4\inn \TM_4$ on the %%@
base manifold, as described in equations~\ref{hvvv} and \ref{hinhtho}.
 These features clearly require further investigation.

 Within the above caveats, aligned with the charge magnitudes of 1 and $\frac{1}{3}$ %%@
for the electron and $d$-quark Weyl spinors  of equation~\ref{suthuoth} the respective %%@
$\uo_Q$ charges of 
  $0$ and $\frac{2}{3}$ in  equation~\ref{suthuoa} correlate  with
 charges of $\binom{\; 0}{-1}$ for the $\binom{\nu}{e}$ lepton doublet and %%@
$\binom{+2/3}{-1/3}$ for the $\binom{u}{d}$ quark doublet of the Standard Model. 
 In addition the states associated with each \textit{left-handed} doublet of charges %%@
interact via the exchange of $W^{\pm}$ gauge bosons in the Standard Model.
 Hence it remains to be understood how interactions within each of these doublets may %%@
be mediated via an $\sutw_L$ symmetry, and how such $\nu$-lepton and $u$-quark %%@
components of $a\inn \ooo \subset \htho$ gain a Weyl spinor structure under the %%@
external $\sltc^1$ action. 
Here we initially focus upon the simple fact that weak $\sutw_L$ transformations act %%@
on fermion doublets of the form $\binom{\nu}{e}$ and $\binom{u}{d}$, which have been %%@
associated with the $\binom{a}{\theta}$ components of $\htho$ in equation~\ref{xoct3} %%@
for the present theory.

The type 1 $\sltc^1$ action on 
  the four Weyl spinors of equation~\ref{thcth234}
  is  complemented by $\sltc^2$ and $\sltc^3$ transformations of type 2 and 3, all %%@
involving quaternion algebra composition with $l\inn \ooo$ being the only imaginary %%@
octonion unit appearing in the transformation matrices.
 Two SU(2)s are  immediately identifiable in terms of the rotation subgroups of the %%@
type 2 and type 3 Lorentz groups, as denoted by $\sutw^2$, generated by the set %%@
$\{\dot{R}_{z \pqg l}^{2},\dot{R}_{x \pqg z}^{2}, \dot{R}_{x \pqg l}^{2}\} $, and %%@
$\sutw^3$, as generated by $\{\dot{R}_{z \pqg l}^{3},\dot{R}_{x \pqg z}^{3}, %%@
\dot{R}_{x \pqg l}^{3} \}$.
Neither $\sutw^2 \subset \sltc^2$ nor $\sutw^3 \subset \sltc^3$ is independent of %%@
$\sltc^1$ within the $\esi$ Lie algebra, with for
example $\lbrack \dot{R}_{x \pqg z}^{2}, \dot{R}_{x \pqg z}^{1} \rbrack = \fhs %%@
\dot{R}_{x \pqg z}^{3} \neq 0$, and neither of them  forms  a subgroup of   
$\stab$, and hence they do \textit{not}  form an \textit{internal} symmetry by the %%@
original definition which led to table~\ref{stabset}.
 However owing to the properties described below in exploring further the structure of %%@
these transformations the groups $\sutw^{2,3}$ \textit{are} found to be of some %%@
interest in relation to the structure of electroweak theory.

    The type 1 action of $\sltc^1$  decomposes the space $\theta^1 = %%@
\binom{c}{\bar{b}} \inn \ooo^2$ into the four Weyl spinors of equation~\ref{thcth234}. %%@
The transformations ${\sltc^{2,3}}$ of type 2 and 3, with complementary transformation %%@
matrices also based on the units $\{1,l\}$, similarly respect the octonion %%@
decomposition aligned to the four base unit sets: 
\begin{equation}
   \{1,l\}, \quad \{\il,i\}, \quad \{\jl,j\}, \quad \{\kl,k\}
   \label{lijksets}
\end{equation}
 based on the same quarternion subalgebras, now for all three of $a,b,c\inn \ooo$. 
  Hence  the subgroups $\sutw^{2,3} \subset \esi$ describe transformations between the  %%@
components of equation~\ref{suthuoth} and those of equation~\ref{suthuoa} respecting %%@
the alignment of the four component pieces, and hence acting independently on the %%@
corresponding doublets of leptonic and quark states as appropriate for weak %%@
interactions. 
 With respect to the embedding of $a,b,c \inn \ooo$ as components of $\htho$ in %%@
equation~\ref{xoct3}, the spinor representation mixing actions of $\sltc^{1,2,3}$  can %%@
also be displayed graphically as: 
\begin{equation}
  \left(
  \setlength{\unitlength}{10pt}
	 \begin{picture}(9.6,5.6)(0,0)
	     \put(4.6,4.5){$\bar{a}$}
		 \put(9.1,4.5){$c$}
		 \put(0.1,0){$a$}
		 \put(9.1,0){$\bar{b}$}
		 \put(0.1,-4.5){$\bar{c}$}
		 \put(4.6,-4.5){$b$} 
	    \thicklines 
		 \put(5.6,4.8){\vector(-1,0){0.1}}
		 \put(8.7,4.8){\vector(1,0){0.1}}
		  \multiput(5.9,4.8)(1,0){3}{\line(1,0){0.5}}
		 \put(1.1,0.3){\vector(-1,0){0.1}}
		 \put(8.7,0.3){\vector(1,0){0.1}}
	      \multiput(1.4,0.25)(0.25,0){28}{$.$}
		 \put(1.1,-4.2){\vector(-1,0){0.1}}
		 \put(4.3,-4.2){\vector(1,0){0.1}}
		  \put(1.1,-4.2){\line(1,0){3.2}}		
		 \put(9.4,4.1){\vector(0,1){0.1}}	 
		 \put(9.4,1.4){\vector(0,-1){0.1}}
		  \put(9.4,1.4){\line(0,1){2.7}}
		 \put(0.4,-0.4){\vector(0,1){0.1}}
		 \put(0.4,-3.5){\vector(0,-1){0.1}}
		  \multiput(0.4,-3.3)(0,1){3}{\line(0,1){0.5}}
		 \put(4.9,4.1){\vector(0,1){0.1}}
		 \put(4.9,-3.4){\vector(0,-1){0.1}}	
		  \multiput(4.75,-3.0)(0,0.25){28}{$.$}
	 \end{picture}
   \right)
    \qquad \mbox{with} \qquad
	 \begin{array}{ll}
	    \setlength{\unitlength}{10pt}
	    \begin{picture}(3,1) 
		 \thicklines    
		 \put(0,0.5){\line(1,0){3}}
		 \put(0,0.5){\vector(-1,0){0.1}}
		 \put(3,0.5){\vector(1,0){0.1}}
	    \end{picture}
	   &  \sltc^1   \\
	    \setlength{\unitlength}{10pt}
	    \begin{picture}(3,1) 
		 \thicklines    
		 \multiput(0.25,0.5)(1,0){3}{\line(1,0){0.5}}
		 \put(0,0.5){\vector(-1,0){0.1}}
		 \put(3,0.5){\vector(1,0){0.1}}
	    \end{picture}
	   &  \sltc^2   \\ 
	   \setlength{\unitlength}{10pt}
	    \begin{picture}(3,1) 
		 \thicklines    
		 \multiput(0.35,0.5)(0.25,0){9}{$.$}
		 \put(0,0.5){\vector(-1,0){0.1}}
		 \put(3,0.5){\vector(1,0){0.1}}
	    \end{picture}
	   &  \sltc^3  
	     \end{array}
		  \label{abcmix}
\end{equation}

 This indicates how the $\binom{c}{\bar{b}}$ spinor components under $\sltc^1$ are %%@
replaced by $\binom{a}{\bar{c}}$ and $\binom{b}{\bar{a}}$ spinors under $\sltc^2$ and %%@
$\sltc^3$ respectively, depending on the alignment of the %%@
$\theta^a=\binom{\theta_1}{\theta_2}$ components, with $\theta^1$ depicted in %%@
equations~\ref{xoct3}. 
 It is the observation that the $\sltc^{2,3}$ actions relate the $\theta^1 = %%@
\binom{c}{\bar{b}} \inn \ooo^2$ components with the $a\inn \ooo$ component in %%@
equations~\ref{abcmix}, while respecting the four-way octonion decomposition of %%@
equation~\ref{lijksets}, that suggests that these transformations might be closely %%@
related to the weak interactions.

  The nine generators of the combined type $a=1,2$ and 3 rotations $\sutw^a \subset %%@
\sltc^a$ form a closed subalgebra of $\esi$, which is  eight dimensional due to the %%@
linear dependence of the three $\dot{R}^a_{x \pqg l}$ generators. This subalgebra is %%@
in fact an $\sutha$, a linearly independent basis for which can be described by the %%@
eight rotation generators (\cite{Wang} p.128):
\begin{equation}
  \label{rrrrrrrr}
    \sutha_s \, \equiv \,
     \{ \dot{R}^1_{x \pqg l},\,  \dot{R}^2_{x \pqg l},\,  \dot{R}^1_{x \pqg z},\, 
   \dot{R}^2_{x \pqg z},\,  \dot{R}^3_{x \pqg z},\,  \dot{R}^1_{z \pqg l}, \, 
   \dot{R}^2_{z \pqg l},\,  \dot{R}^3_{z \pqg l}  \}  
\end{equation}
  These generate a group denoted $\suth_s$ (where `$s$' denotes the `standard' %%@
representation   or embedding of $\suth$ in $\esi$ \cite{Wang}). 
 Since each $\sltc^a$ is independent of $\suth_c$,  
within $\esi$ the subgroup $\suth_s$ is also independent of the colour subgroup %%@
$\suth_c$, as generated by the eight elements on the left-hand side of  %%@
table~\ref{suttan}, with the Lie bracket composition of any element of %%@
equation~\ref{rrrrrrrr} with any element of $\{\dot{A}_q, \dot{G}_l\}$ being zero.

  Guided by a similar construction based on $\sutw_L$ generators in the Standard %%@
Model, here in the complex algebra for the generators of the subgroup $\sutw^2 \subset %%@
\esi$ we define: 
\begin{equation}
  \label{sigcomb}
   \dot{\Sigma}^{(2)\pm} := \, \dot{R}_{z \pqg l}^2 \; \pm \; i\dot{R}_{x \pqg z}^2
\end{equation} 
  Here the imaginary unit $i\inn \ccc$ in the complexification of the $\esi$ Lie %%@
algebra  commutes with the elements of  $T\htho$, which are based on an independent %%@
octonion algebra $\ooo$. 
 On  reading off the corresponding entries in the Lie algebra table in~\cite{Wang} for %%@
the complex element of equation~\ref{sigcomb} it is found that:
\begin{eqnarray}
    \lbrack  \dot{S}^1_l \, , \, (\dot{R}_{z \pqg l}^2 \, + \, i\dot{R}_{x \pqg z}^2)
	\rbrack & = &  \mbox{\small $\frac{3}{2}$} \dot{R}_{x \pqg z}^2 \; - \; i 
	         \mbox{\small $\frac{3}{2}$}    \dot{R}_{z \pqg l}^2    \; = \;
	-i\mbox{\small $\frac{3}{2}$} 
	   (\dot{R}_{z \pqg l}^2 \, + \, i\dot{R}_{x \pqg z}^2) \label{srrbrac} \\
 \!\!\!\!\!\!\!\!\!\!\!	\mbox{hence} \qquad 
	 \lbrack  i\Sbard^1_l \, , \, (\dot{R}_{z \pqg l}^2 \, + \, i\dot{R}_{x \pqg z}^2)
	\rbrack & = &   +  (\dot{R}_{z \pqg l}^2 \, + \, i\dot{R}_{x \pqg z}^2) \nonumber %%@
\\
 \!\!\!\!\!\!\!\!\!\!\! \mbox{and} \qquad \qquad \quad \;\;
	 \lbrack  i\Sbard^1_l \, , \,  \dot{\Sigma}^{(2)\pm} 
	\rbrack     & = &   \pm \dot{\Sigma}^{(2)\pm}    \label{isscomm}
\end{eqnarray} 
  with real charge eigenvalues $\pm 1$.
  Hence the generators $\dot{\Sigma}^{(2)\pm}$ of equation~\ref{sigcomb} are %%@
associated with the \textit{same} magnitude of $\uo_Q$ charge under $\Sbard^1_l$ as %%@
was found for the electron in the leptonic components $\theta_l \subset \htho$ as %%@
described in equations~\ref{sloncb}--\ref{suthuoth}.
  The generators $\dot{\Sigma}^{(2)\pm}$ hence act as charge raising and lowering %%@
operators, analogous to the action of $W^{\pm}$ gauge bosons in the Standard Model, %%@
here deriving from the embedding of $\sutw^2 \subset \suth_s$. 

   Further, the $\esi$ generator linear dependences described in $\cite{Wang}$ imply %%@
the relation:
\begin{equation}
  \label{srslincom2}
   - \, \Sbard^1_l = \dot{R}_{x \pqg l}^2 + \fhs \Sbard^2_l 
\end{equation}
   which, within a choice of sign conventions, is closely reminiscent of the Standard %%@
Model relation:
\begin{equation}
  \label{qethy2}
    Q \, = \,  T^3 + \fhs Y   
\end{equation}
	between the charge $Q$, the third $\sutw_L$ generator $T^3$ and hypercharge $Y$.
	This suggests associating $ \Sbard^2_l$ with $ Y$ as a candidate for the generator %%@
of the
  \textit{hypercharge}  symmetry $\uo^2 \sim \uo_Y$ which commutes with $\sutw^2$, as %%@
generated by 
 $\{\dot{R}_{z \pqg l}^2, \dot{R}_{x \pqg z}^2, \dot{R}_{x \pqg l}^2  \}$.

 The $\sutw^2 \times \uo^2$ symmetry  generated by $\{ \dot{R}_{z \pqg l}^2, %%@
\dot{R}_{x \pqg z}^2,
  \dot{R}_{x \pqg l}^2, \Sbard^2_l\}$
   acts on the doublet components of the type 2 spinor $\theta^{2} = %%@
\binom{a}{\bar{c}}$ in $\htho$. Restricted to the complex subspace $\ccc\subset \ooo$ %%@
with $\{1,l\}$ basis units the components   $\theta^{2}_l = \binom{a}{\bar{c}}_{\! l}$  %%@
provisionally represents the lepton doublet $\binom{\nu}{e}$.  
Since these components do not correspond to complete $\sltc^1$ Weyl spinors for either %%@
the neutrino \textit{or}  the electron part this $\sutw^2 \times \uo^2$ symmetry is %%@
clearly not directly equivalent to the $\sutw_L \times \uo_Y$ symmetry of electroweak %%@
theory. However 
the components of $\theta^{2}_l$ do transform under the internal symmetry $\suth_c %%@
\times \uo_Q$  appropriately to represent such a lepton doublet, as described earlier %%@
in this section, and hence the $\sutw^2 \times \uo^2$ symmetry
  serves as a useful intermediate model, considered as a `mock electroweak theory'.

 It is also the case that  neither the $\sutw^2$ generated by $\{ \dot{R}_{z \pqg %%@
l}^2, \dot{R}_{x \pqg z}^2,
  \dot{R}_{x \pqg l}^2 \}$ nor the  $\uo^2$ generated by $\Sbard^2_l$  are internal %%@
symmetries in the sense of table~\ref{stabset}, that is within $\stab$, with each of %%@
these four generators impacting upon the components of the type 1 subspace $\htwc %%@
\subset \htho$, which represent components of the external spacetime $\TM_4$, as can %%@
be seen explicitly from the form of these four generators in tables~\ref{lbrota} and %%@
\ref{ltrota}. The breaking of the full $\esi$ symmetry action on $\htho$ in this %%@
identification of the type 1 subspace $\htwc$ with the local tangent space $\TM_4$ of %%@
the external spacetime hence includes the breaking of the $\sutw^2 \times \uo^2 %%@
\subset \esi$ subgroup.

  With the one generator surviving this $\sutw^2 \times \uo^2$ symmetry breaking being %%@
the $\Sbard^1_l$  generator on the left-hand side of equation~\ref{srslincom2}, which %%@
has been associated with the $\uo_Q$ of electromagnetism, this structure is closely %%@
analogous to electroweak theory in the Standard Model. This motivates the proposal %%@
that the four components of $\bv_4 \inn \TM_4 \equiv \htwc \subset \htho$ might be 
 considered to form a `vector-Higgs', in place of the four scalar components of the %%@
Standard Model Higgs field, underlying empirically observed Higgs phenomena. The %%@
scalar magnitude $\vert \bv_4 \vert$, or another scalar deriving from the $\bv_4$ %%@
components might then constitute the Higgs scalar itself, in a manner resembling %%@
composite Higgs models. Mass terms for gauge fields associated with the broken 
 $\sutw^2 \times \uo^2$ generators would then derive from the fact that they do not %%@
generate elements of $\stab$, while fermion masses originate from interaction terms %%@
implied in the composition of $\lvt$ in equation~\ref{detpmn}, or a higher-dimensional %%@
extension.

  Using the  $\esi$ Lie algebra table~\cite{Wang} a basis for the $\sutha_s$ Lie %%@
subalgebra can be obtained with a normalised  
   Killing form with  components proportional to the unit $8 \times 8$ matrix.
 This can be achieved by replacing the first two basis elements in %%@
equation~\ref{rrrrrrrr} by $\{\dot{R}^a_{x \pqg l}, \, \frac{1}{\sqrt{3}}\dot{S}^a_l %%@
\}$ for three possible choices corresponding to  type $a= 1,2,3$.
 Using this normalisation the value of the mixing angle $\thetmt$ in the breaking of %%@
the $\sutw^2 \times \uo^2$ symmetry to $\uo_Q$ may be determined, by analogy with the %%@
weak mixing angle $\theta_W$ in standard electroweak theory.

 A significant difference is observed between the calculated value  $\sin^2 \thetmt %%@
=\frac{3}{4}$
  and the known empirical 
 $\sin^2 \theta_W \simeq 0.23$.
 However, as emphasised earlier in this subsection the $\sutw^2 \times \uo^2$ symmetry %%@
does \textit{not} act on $\sltc^1$ Weyl spinors in the appropriate way to describe %%@
weak interactions, and here we are dealing with a provisional `mock electroweak %%@
theory', which nevertheless exhibits some of the features associated with 
corresponding structures of the Standard Model, such as the \textit{possibility} of
% possibility of identifying  a mixing angle.
 identifying a mixing angle.
  It is also noted that in the mock theory the calculated value of $\sin^2 \thetmt %%@
=\frac{3}{4}$ effectively corresponds to a `unification scale' whereas the empirical %%@
value of 
 $\sin^2 \theta_W \simeq 0.23$ is determined at the practical energy scale of $M_Z %%@
\sim 10^{2}\,$GeV.

   However while the properties of $\sutw^2 \times \uo^2$ described above are %%@
reminiscent of several features of electroweak group it is well known that in fact the %%@
full rank-6 group $\sltc \times \suth \times \sutw \times \uo$, representing the %%@
external spacetime symmetry and the internal symmetry of the Standard Model, cannot be %%@
accommodated as a subgroup within $\esi$, for example by the analysis of Dynkin %%@
diagrams. This observation, and the need to identify further features of the Standard %%@
Model fermions, suggests consideration of a larger symmetry for a higher-dimensional %%@
form of time $\lvh$ beyond the 27-dimensional cubic form considered in this section.

\pagebreak

\section{$\ese$ Symmetry and Left-Right Asymmetry}
\label{secesef}

   A higher-dimensional  \textit{homogeneous} polynomial form  is desired, in %%@
conformity with the discussion of equation~\ref{dsxxx} and more generally with the %%@
derivation of equation~\ref{lv} in section~\ref{sym}. While the determinant preserving %%@
symmetry of the space $\mcX \inn \htho$ describes the lowest-dimensional non-trivial %%@
representation of $\esi$ the smallest non-trivial representation of the exceptional %%@
Lie group E$_7$ is 56-dimensional and may be constructed in terms the elements $x$ of %%@
the Freudenthal triple system $F(\htho)$ (\cite{Krute,Borst,Rios}, \cite{Baez1} p.48).

  The vector space $F(\htho)$ has 56 real components and may be introduced according %%@
to Freudenthal's construction with the vector space composition (which may be compared %%@
with the further decomposition of equation~\ref{horhoo}):
\begin{equation}
  \label{fvecspa}
  F(\htho) \; \cong \; \htho \, \oplus \, \htho  \, \oplus \, \rrr \, \oplus \, \rrr
\end{equation}
  Correspondingly elements $x \inn F$, with $F=F(\htho),$ can generally be written in %%@
the form of a `$2 \times 2$ matrix' as:
\begin{equation}
  \label{ftscomp}
   x =  \left(\begin{array}{cc} \alpha & \mcX \\
                          \mcY & \beta \end{array} \right), \qquad
		\mbox{with} \;\;   \mcX,\mcY \inn \htho,
						   \quad \alpha,\beta \inn \rrr
\end{equation}
\begin{equation}
 \label{hthoxy}
 \mbox{and} \qquad \quad  \mcX = 
   \left( \begin{array}{ccc}
       p & \bar{a} & c  \\
       a &   m     & \bar{b}        \\
 \bar{c} &   b     & n 
          \end{array}  \right), \qquad 
   \mcY = 
   \left( \begin{array}{ccc}
       P & \bar{A} & C  \\
       A &   M     & \bar{B}        \\
 \bar{C} &   B     & N 
          \end{array}  \right)
\end{equation}
  here with the real $P,M,N$ and octonion $A,B,C$ components of $\mcY$ distinguished %%@
from the lower case counterpart components of $\mcX$. 
 A homogeneous quartic norm $q:F \to \rrr$ can be defined on the components of $x\inn %%@
F$ as follows:
\begin{equation}
   q(x) \; = \; -2\lbrack \alpha\beta - (\mcX,\mcY)\rbrack^2 \, - \,
       8\lbrack\alpha \det(\mcX) + \beta\det(\mcY) - (\mcX^{\sharp},
	                                          \mcY^{\sharp})\rbrack \label{fquartic}									  			 
\end{equation}
   where all the necessary definitions contained within this expression are inherited %%@
from those for the Jordan algebra $\htho$ as described in \cite{Krute,Borst,Rios}.
   The group Inv($F$) of all invertible transformations $\sigma$ in $F$ preserving the %%@
above quartic norm with $q(\sigma(x)) = q(x)$
 is found to be the non-compact real form E$_{7(-25)}$ of the exceptional Lie group %%@
E$_7$. Hence, in particular, under this symmetry group the invariance of the quartic %%@
form $q(x)$, as a homogeneous polynomial, describes a possible 56-dimensional form of %%@
temporal flow in the form of equation~\ref{lv} which may be denoted $L(\bv_{56})=1$. 
  This provides a natural extension from the cubic form $\lvt$ with $\esi$ symmetry %%@
described in the previous two sections.

 Including the actions of the subgroup $\esi \subset \mbox{E}_7$  on the elements  
$x \to s(x) \inn F$ (here $s$ and $x$ conform with the notation in %%@
references~\cite{Krute,Borst,Rios}, their meaning is of course very different to that %%@
of $s$ and $x$ in equation~\ref{dsxxx} for example) the transformations of the full %%@
symmetry $\ese \equiv \mbox{Inv}(F)$ may be categorised in terms of  four sets. With %%@
$s \inn \esi$, $\lambda \inn \rrr$ and $C,D \inn \htho$ these %%@
are~\cite{Krute,Borst,Rios}: 
\begin{eqnarray}
 \!\!\! T(s): \,  \left(\begin{array}{cc} \alpha & \mcX \\
                          \mcY & \beta \end{array} \right) & \!\! \to \!\! &
		  \left(\begin{array}{cc} \alpha & s(\mcX) \\
                          {s^{\ast}}^{-1}(\mcY) & \beta \end{array} \right)		
						   \label{ftsstran}		   \\ 
 \!\!\! \lambda: \,   \left(\begin{array}{cc}\alpha & \mcX \\
                          \mcY & \beta \end{array} \right) & \!\! \to \!\! &
  \left(\begin{array}{cc}\;\! \lambda^{-1}\alpha\; &\; \lambda^{\frac{1}{3}}\mcX \;\! %%@
\\
                \;\! \lambda^{-\frac{1}{3}}\mcY \;& \; \lambda \,\beta \end{array}\;\!
				   \right)	\label{lamtran} \\
\!\!\! \phi(C): \,   \left(\begin{array}{cc} \alpha & \mcX \\
                          \mcY & \beta \end{array} \right) & \!\! \to \!\! &
		  \left(\begin{array}{cc}
 \!  \alpha+(\mcY,C) + (\mcX,C^{\sharp}) + \beta \det(C) \, & \, \mcX + \beta C \! \\
        \!  \mcY + \mcX\times C + \beta C^{\sharp} \, & \, \beta \end{array} \! %%@
\right)		
						\label{phicact}     \\ 
\!\!\!  \psi(D): \, \left(\begin{array}{cc} \alpha & \mcX \\
                          \mcY & \beta \end{array} \right) & \!\! \to \!\! &
		  \left(\begin{array}{cc}
  \! \alpha \, & \, \mcX + \mcY\times D + \alpha D^{\sharp} \! \\
  \! \mcY + \alpha D \, & \,
      \beta + (\mcX,D) + (\mcY,D^{\sharp}) + \alpha \det(D) \end{array} \! \right)
	  \quad
	    \label{psidact}	
\end{eqnarray}   
   The set of actions ${s^{\ast}}^{-1}$ in equation~\ref{ftsstran} is equivalent to %%@
the complex conjugate of the representation defined by the set of actions $s\inn \esi$ %%@
on $\htho$. Under the subgroup $\esig \subset {\mbox{E}_{7(-25)}}$ the space $F$ %%@
decomposes into the reducible representation (\cite{Borst} equations~9.45 and 9.46):
\begin{equation}
  \label{esetoesi}
    \mathbf{56}_{\mathrm{E}_7} \, \to \, 
	(\mathbf{27} + \mathbf{\overline{27}} + \mathbf{1} + \mathbf{1})_{\mathrm{E}_6}
\end{equation}	
  compatible with the structure of equation~\ref{fvecspa}
  (and can be compared with the further reduction under $\spotn$ in %%@
equation~\ref{esidecom}).

 The four sets of group actions of equations~\ref{ftsstran}--\ref{psidact} are %%@
described at the Lie algebra level in reference~\cite{DrayMW}. In particular it is %%@
shown that for the first set of  equation~\ref{ftsstran} the $\esi$ `rotations' are %%@
identical on $\mcX$ and $\mcY$ while for the $\esi$ `boosts' there is a change in sign %%@
between the actions on the two $\htho$ subspaces.
   Having extended beyond the $\esi$ subalgebra to the full $\ese$ we next focus on %%@
the generators of the 4-dimensional spacetime Lorentz subgroup $\sltc^1 \subset \esi %%@
\subset \ese$ of type 1 as studied in the first part of the previous section. 
  For these Lorentz actions
 reversing the sign of the boosts is precisely the operation which interchanges %%@
between the left-handed and right-handed spinor representations of $\sltc$.

  Hence while the components of $\theta_l$ in $\theta^1$ within  $\mcX  \inn \htho$, %%@
defined in equation~\ref{thcth234}, transform as a \textit{left}-handed Weyl spinor %%@
under $\sltc^1$ the corresponding components of $\theta_{\!\lag} = 
\binom{C_1 + C_8l}{B_1 - B_8l}$ within the $\theta^1$ component of $\mcY \inn \htho$, %%@
extracted from equation~\ref{hthoxy}, transform as a \textit{right}-handed  Weyl %%@
spinor under the same $\sltc^1 \subset \esi \subset \ese$ action.
 (The subscript `$\lag$' on $\theta_{\!\lag}$ denotes both the use of the imaginary %%@
unit $l$ and the identification of the `leptonic' components of $\theta^1 = %%@
\binom{C}{\bar{B}}$ in $\mcY$, as will be seen below. In general the type superscript %%@
`1' is not appended to components such as $\theta_l$ and $\theta_{\!\lag}$ since they %%@
are unambiguously extracted from  `type 1' $\theta^1$ components,
  while a superscript is included for the `type 2' or `type 3' case).
 Considered as an action of  $2 \times 2$ matrices $S\inn \sltc^1$  on the 2-component %%@
Weyl spinors $\theta_l$ and $\theta_{\!\lag}$, extracted from the corresponding %%@
$\theta^1$ components of $\mcX$ and $\mcY$
respectively the action of equation~\ref{ftsstran} may be summarised as:
\begin{equation}
  \label{diracss}
   \left(\begin{array}{c}  \theta_l \\ \theta_{\!\lag} \end{array} \right)
			 \; \to \;			  
		 \left(\begin{array}{cc} S  &  0  \\
                          0 & {S^{\dag}}^{-1} \end{array} \right)
         \left(\begin{array}{c} \theta_l \\ \theta_{\!\lag}  \end{array} \right)
\end{equation}
  This is precisely the Lorentz transformation of a 4-component Dirac spinor $\psi$.

 As explained in section~\ref{chapesb} the components of $\theta^1$ within $\mcX \inn %%@
\htho$ under the action of $\sltc^1$ actually decompose into a set of four left-handed %%@
Weyl spinors
 $\{\theta_l, \theta_i, \theta_j, \theta_k\}$ as listed in equation~\ref{thcth234}. 
 Hence equation~\ref{ftsstran} contains both the original representation of $\sltc^1$ %%@
on $\mcX$, which contains the set of four left-handed Weyl spinors in the $\theta^1$ %%@
components, simultaneously with an equivalent of the complex conjugate representation %%@
on $\mcY$, which hence contains a corresponding set of four right-handed Weyl spinors, %%@
which may be denoted $\{\theta_{\! \lag}, \theta_I, \theta_J, \theta_K\} \subset %%@
\mcY$. Correspondingly a set of four 4-component Dirac spinors have hence been %%@
identified with:
\begin{equation}
 \label{diraclr}
   \psi = \left(\begin{array}{c}  \psi_L \\ \psi_R \end{array} \right) \; = \;
     \left(\begin{array}{c}  \theta_l \\ \theta_{\!\lag} \end{array} \right), \quad
	 \left(\begin{array}{c}  \theta_i \\ \theta_I \end{array} \right), \quad
	 \left(\begin{array}{c}  \theta_j \\ \theta_J \end{array} \right)
	 \quad \mbox{or} \quad
	 \left(\begin{array}{c}  \theta_k \\ \theta_K \end{array} \right) 
\end{equation}
   each of which transforms in the manner of equation~\ref{diracss}
  under $S\inn \sltc^1 \subset \esi \subset \ese$ transformations.

  The internal $\suth_c \times \uo_Q$ symmetry, also described in the previous %%@
section, is composed as a subgroup of $\esi$ purely out of the subset of rotations. %%@
Hence these actions are identical on the components of $\mcX$ and $\mcY$ in %%@
equation~\ref{ftsstran}. Hence in turn the $\suth_c$ action on the components of %%@
$\mcX$, including upon the $\theta^1$ components as detailed in table~\ref{suttan} and %%@
summarised together with the $\uo_Q$ action in equation~\ref{suthuoth}, is identical %%@
for the corresponding components of $\mcY$, and the corresponding $\uo_Q$  charges for %%@
the respective subcomponents of equation~\ref{hthoxy} are also the same. Hence the %%@
$\psi_L$ and $\psi_R$ components carry matching $\suth_c \times \uo_Q$ transformation %%@
properties for the set of four Dirac spinors in equation~\ref{diraclr}
(justifying the identification of both $\theta_l$ and $\theta_{\!\lag}$ as leptonic
  components). 
  Similarly the $\sutw^{2,3} \times \uo^{2,3} \subset \esi$ rotations, for the  mock %%@
electroweak theory described in section~\ref{chapesb}, also act on the $\mcX$ and %%@
$\mcY$ components of $x \inn F(\htho)$ in the same way.

  While the total number of dimensions has been increased from 27 to 56 it remains the %%@
case that only a single set of 4 dimensions will describe the external spacetime. This %%@
can be chosen as an $\htwc \subset \htho$ subset of components  $\bv_4$ within $\mcX$, %%@
under an $\sltc \subset \esi$ action, \textit{or} as an $\htwc \subset \htho$ subset %%@
of components  $\bv_4$ within $\mcY$, transforming under the complex conjugate %%@
representation, but not \textit{both}. Here we choose 
  $\bv_4 \equiv \bh_2 \inn \htwc$ as embedded within the $Y=\binom{P \; \bar{A}}{A \; %%@
M} \inn \htwo$ components of type~1 in $\mcY$ in equation~\ref{hthoxy} to represent %%@
external spacetime,
 with Lorentz transformations hence described by:
\begin{equation}
\label{hshsdual}
  \bh_2 \to  \bh_2^{\prime} =  {S^{\dag}}^{-1} \, \bh_2 \, S^{-1}
\end{equation}
  rather than equation~\ref{hshs}, under the action of $S\inn \sltc^1 \subset \esi$.
 The complex subspace with base units $\{1,l\}$ still underlies 
 both the $\sltc^1$ subgroup and the subspace for the vectors $\bh_2 \inn \htwc$.    
 These $\bh_2$ components of $\mcY$ will also now be taken to
  form the `vector-Higgs' 
   correlated with the phenomena of the Standard Model Higgs sector and  Yukawa %%@
couplings, as was described for the original case of $\lvt$ towards the end of the %%@
previous section. Here for the case of $\lvfs$ this now implies that \textit{none} of %%@
the 27 components of $\mcX \inn \htho \subset F(\htho)$ are identified with components %%@
of the external spacetime vectors $\bv_4 \inn \TM_4$.

In particular this means that in addition to the $d$-quark and charged lepton %%@
components of left-handed Weyl spinors in $\theta^1 \subset \mcX$ as identified for %%@
equation~\ref{suthuoth}, potentially both $u$-quark \textit{and} neutral lepton %%@
left-handed Weyl spinors might be identified in the $X$ components of $\mcX$ as %%@
described provisionally in the previous section. The $a\inn \ooo$ component of $\mcX$ %%@
has the correct $(0,\frac{2}{3})$ charge structure to describe ($\nu$-lepton, %%@
$u$-quark) particle states, as seen in equations~\ref{sbcharge} and \ref{suthuoa}, and %%@
is now free to accommodate both states.
 However while the corresponding imaginary $A(6)$ components of $\mcY$ also have an %%@
$\Sbard_l^1$ charge of $\frac{2}{3}$, the $A_{1,l} = (A_1 + A_8l)$ part of $A\inn %%@
\ooo$ in $\mcY$ is \textit{occupied} by the above components $\bh_2 \inn \htwc$, %%@
representing the vector-Higgs and external spacetime, 
 as depicted in equation~\ref{fhthopart}.
\begin{equation}
 \hspace*{-20pt}
  \label{fhthopart}
  \left(\!\!\!\!\! \begin{array}{cc}   \alpha &   \!\!\!
     \left(\!\! \begin{array}{c|c} 
	      \, X \!\!\sim\thX\thX^{\dag} 
		  		   \!\!\!\!\! \begin{array}{cc} &  \\  &  \end{array} \!\!\!   &
        \,  \theta^1  \begin{array}{cc} &  \\  &  \end{array} \!\!\!\!\!\!\!\!\!\! 
				                         \\  \hline
        \,\,\,\,\,\,\,\, {\theta^1}^{\dagger} \!\!\!\! \begin{array}{cc}  
		      &   \end{array}   &	  
		\,  n      \end{array}  \!\! \right)_{\mbox{$\!\!\!\!\!\;\mcX$}}   \\
		\left(\!\!  \begin{array}{c|c} 	
	\, Y \!\!\sim\thY\thY^{\dag}
	  \!\!\!\!\! \begin{array}{cc} &  \\  &  \end{array} \!\!\!   &		  
        \,  \theta^1  \begin{array}{cc} &  \\  &  \end{array} \!\!\!\!\!\!\!\!\!\! 
				                         \\  \hline
        \,\,\,\,\,\,\,\, {\theta^1}^{\dagger} \!\!\!\! \begin{array}{cc} 
		         &   \end{array}   &	  
		\,  N      \end{array} \!\! \right)_{\mbox{$\!\!\!\mcY$}}
             \!\!\!          & \beta  \end{array} \!\!\!\!\!\right)
\quad \sim \quad
  	 \left(\!\!\!\!\! \begin{array}{cc}    &   \!\!\!
     \left(\!\! \begin{array}{c|c} 
         \begin{array}{c} \;\;\, \mbox{`}\nu_L\mbox{'} \;\;\,  \\
		   \mbox{`}u_L\mbox{'}  \end{array}    &
          \!\!  \begin{array}{c}    e_L   \\ d_L    \end{array} \!\! 
				                         \\  \hline
        \begin{array}{cc}  
		           &   \end{array}   &	  
		          \end{array}  \!\! \right)_{\mbox{$\!\!\!\!\!\;\mcX$}}   \\
		\left(\!\! \begin{array}{c|c} 
         \begin{array}{c} \! \bv_4 \equiv \bh_2  \!  \\ 
		     \;\;    \mbox{`}u_R\mbox{'} \;\;  \end{array}    &
          \!\!  \begin{array}{c}    e_R   \\ d_R    \end{array} \!\! 
				                         \\  \hline
        \begin{array}{cc}  
		           &   \end{array}   &	  
		          \end{array}  \!\! \right)_{\mbox{$\!\!\!\mcY$}}
             \!\!\!          &  \end{array} \!\!\!\!\!\right) 
\end{equation}

This in principle provides an explanation for the existence of the left-handed %%@
neutrino $\nu_L$ while the corresponding right-handed state $\nu_R$ is prohibited, at %%@
least at the level of the basic symmetry structures, as a feature of the breakdown of %%@
left-right symmetry through the identification of external spacetime in the breaking %%@
of the full symmetry of  $\lvfs$.
 This observation is accompanied by the caveat concerning the Weyl spinor composition %%@
of the components of $X\subset \mcX$ and $Y \subset \mcY$.
 With this in mind, and hence with quote marks  placed on the $\nu_L$, $u_L$ and $u_R$ %%@
states, the relation between the component 
 structure for elements of $x \inn F(\htho)$, in the form of equations~\ref{ftscomp} %%@
and \ref{hthoxy}, and the first generation of Standard Model fermions is summarised in 
  equation~\ref{fhthopart}.

   Unlike the case for $\esi$, the Lie algebra $\ese$ does contain a rank-6 subgroup %%@
corresponding to the combined external Lorentz symmetry and internal gauge symmetry of %%@
the Standard Model, that is:
\begin{equation}
 \label{esedecom}
   \sltc \; \times \; \suth \times \sutw \times \uo \;\;\, \subset \;\;\, \ese
\end{equation} 
  Since a similar decomposition cannot be obtained from the $\esi$ symmetry alone, as %%@
described at the end of the previous section, this suggests that the identification of %%@
an internal $\sutw$ symmetry will require the use of some combination of elements from %%@
$\phi(C)$ and $\psi(D)$ in equations~\ref{phicact} and \ref{psidact}. Since  these %%@
additional actions differ on the $\mcX$ and $\mcY$ components an $\sutw$ identified in %%@
this way might be expected to have an asymmetric action on the left and right-handed %%@
spinors identified in equation~\ref{fhthopart}.

   With the aim of identifying  an $\sutw_L \times \uo_Y \subset \ese$ symmetry %%@
action, and in contrast with  table~\ref{stabset} in  section~\ref{chapesb}, an %%@
internal symmetry might be defined as any group $\ul{G}$ consistent with the subgroup
  decomposition $\sltc^1 \times \ul{G} \subset \ese$ for which the set of $\sltc^1$ %%@
spinors transform under the trivial or fundamental representations of $\ul{G}$. That %%@
is, while the external $\sltc^1$ symmetry partitions the components of $\lvh$ into %%@
irreducible pieces, including the spinors $\theta_{l,i,j,k}$ of %%@
equation~\ref{thcth234} and table~\ref{sltcreps} each composed of four real %%@
components, the internal symmetry $\ul{G}$ respects this partitioning  in treating the %%@
Weyl spinors as individual components of a representation of $\ul{G}$.  

The internal $\suth_c \times \uo_Q$ symmetry may also be motivated in this way, as a %%@
component of the $\ese$ decomposition with well-defined representations on the spinor %%@
components, as seen in equation~\ref{suthuoth} for example. In this case the fact that %%@
it also \textit{happens} that $\suth_c \times \uo_Q \subset \stabse \subset \ese$ is %%@
responsible for the fact that the gauge bosons associated with QCD and QED %%@
\textit{happen} to be massless. 
At the same time the action of $\sutw_L \times \uo_Y \subset \ese$  might still be %%@
expected to be closely related to the subgroups $\sutw^{2,3}\times \uo^{2,3} \subset %%@
\esi$ acting on the components of $\mcX$ and $\mcY$, which impinge upon the external %%@
spacetime $\TM_4$ components, since these actions have several suitable properties in %%@
relation to electroweak theory as described for the mock electroweak theory of the %%@
previous section.

 For the present theory 
  the $\sutw_L\times \uo_Y$ symmetry breaks to $\uo_Q$ as generated by $\Sbard^1_l$ of %%@
$\esi \subset \ese$, in the Lie algebra, which acts upon $\mcX$ and $\mcY$ in the same %%@
way. In the case of the $\mcX$ components the $\Sbar^1_l \equiv \uo_Q$ action misses %%@
the $\nu_L$ components of equation~\ref{fhthopart} accounting for the charge %%@
neutrality of the neutrino. In the case of the $\mcY$ components the $\Sbar^1_l \equiv %%@
\uo_Q$ action misses the $\bh_2$ components of equation~\ref{fhthopart} here %%@
potentially accounting for the 
massless nature of the photon.
   These two different aspects of electroweak theory may hence here be described %%@
together in terms of the broken $\ese$ action on the $\mcX$ and $\mcY$ components of %%@
$L(\bv_{56})=q(x)=1$ in equation~\ref{fhthopart}.

   As an extension of the $\lvt$ case described in the previous section, mass terms %%@
for the fermions and proposed to arise from `Yukawa like' couplings between the %%@
fermion and vector-Higgs $\bv_4$ components of equation~\ref{fhthopart}, now through %%@
the quartic constraint $L(\bv_{56})=q(x)=1$ of equation~\ref{fquartic}.

 The scheme in equation~\ref{fhthopart} accounts for one family of quarks and leptons %%@
with the appropriate transformations under the  internal $\suth_c\times \uo_Q$ %%@
symmetry and external $\sltc^1$ symmetry. This observation carries the significant %%@
caveat that the theory requires further development in order to fully identify %%@
$u$-quark and $\nu$-lepton  states that transform appropriately under the external %%@
$\sltc^1$ symmetry. There  are several possible mathematical ways in which to %%@
decompose  the components of $X,Y \inn \htwo$ into a set of spinors under $\sltc^1$,
 such as via the octonion spinor decomposition  $X =  \thX\thX^{\dag}$, or as a sum of %%@
such terms, as suggested on the left-hand side of equation~\ref{fhthopart}. However,
 ideally such a structure should be found to arise naturally within the context of the %%@
present theory rather than introduced arbitrarily.  
In addition  the particle states yet to be identified include the second and third %%@
generation of fermions, as related through CKM mixing in the case of the quark states, %%@
and their relation to the massive gauge bosons associated with electroweak theory in %%@
the Standard Model. In the following section we speculate on the possible nature of a %%@
yet higher-dimensional form of temporal flow in principle capable of accommodating %%@
these phenomena.%%@

\section{$\ee$ Symmetry and the Standard Model}
\label{sosmfi}

  The extension from $\esi$ acting on $\lvt$ to $\ese$ acting on $\lvfs$ can be %%@
considered as a continuation of the progression to higher-dimensional forms of %%@
temporal flow which began with the $\sltc$ Lorentz symmetry of the quadratic form %%@
\mbox{$\lvf$} of equation~\ref{flow4d}, via equation~\ref{vtoh}, on 4-dimensional %%@
spacetime.
   This progression, as an explicit realisation of the underlying motivation for the %%@
present theory as introduced through equations~\ref{dsxx} and \ref{dsxxx}, is %%@
summarised here in table~\ref{lvftolvfs}.
\begin{table}[htbp]
\centering
\begin{tabular}{|l|ccccr|}
 \hline
        &  form & dimensions       & space  & symmetry & \# generators   \\
 \hline
 $\lvf$ & quadratic &  4 spacetime & $\bv_4 \inn \htwc$  & $\sltc$ 
                                                            &  $6 \qquad \;$ \\
 $L(X)\:\!=\:\! 1$ & quadratic &  10 spacetime & $X \inn \htwo$  & $\sltwoo$ 
                                                            &  $45 \qquad \;$ \\
 $L(\mcX)\:\!=\:\! 1$ & cubic & 27 temporal & $\mcX \inn \htho$ & $\mbox{E}_{6(-26)}$ 
                                                            &  $78 \qquad \;$ \\ 
 $L(x)\:= \, 1$ & quartic & 56 temporal & $x \inn F(\htho)$ & $\mbox{E}_{7(-25)}$ 
                                                            &  $133 \qquad \;$ \\ 
   \hline
  \end{tabular}
  \caption{\setb Four-dimensional spacetime, as a form of temporal flow itself, may be %%@
embedded in a progression of higher-dimensional temporal forms.}
\label{lvftolvfs}
\end{table} 
 
 The highest dimensional form of temporal flow $\lvfs$ has a symmetry breaking pattern %%@
to $\mbox{E}_{6(-26)} \subset \mbox{E}_{7(-25)}$  with
 the representations of equation~\ref{esetoesi}
	as exhibited by the structure of equation~\ref{ftsstran}. This is analogous to the %%@
further breaking pattern of $\esi$ to $\sltwoo \equiv \spotn$, as described by the %%@
representations of equations~\ref{horhoo}--\ref{esidecom}, which is also implied in %%@
the structure of left-hand side of equation~\ref{fhthopart}. The $\sltwoo$ symmetry of %%@
10-dimensional spacetime is an intermediate stage on the way down to the Lorentz %%@
$\sltc$ symmetry which further decomposes the representation space into a Lorentz %%@
4-vector, Weyl spinors and Lorentz scalars, as described in table~\ref{sltcreps} and %%@
now applied to both $\mcX$ and $\mcY\inn \htho$, with the external Lorentz 4-vector %%@
$\bv_4 \inn \TM_4$ accommodated within the $\mcY$ components in %%@
equation~\ref{fhthopart}, where two further Lorentz scalar  components $\alpha$ and %%@
$\beta$ are also identified. 

  Apart from the three additional scalars, $N$, $\alpha$ and $\beta$ in %%@
equation~\ref{fhthopart},
   the increase in dimension from 27 to 56 does not contain any redundancy in terms of %%@
comparison with the structures of the Standard Model. Most of the additional 29 %%@
dimensions are interpreted as an augmentation from 2-component  Weyl spinors to %%@
4-component Dirac spinors, together with a separation in the identification of the %%@
left-handed neutrino state and
 the external spacetime $\htwc \equiv \TM_4$ components.

   In terms of the dimension of the underlying space, as listed for the sequence of %%@
forms $\lv$ in table~\ref{lvftolvfs}, we first note that a further expansion from $56$ %%@
to $\sim\,$80
real components would be sufficient to incorporate Weyl spinors for the $\nu_L$, $u_L$ %%@
and $u_R$ states of equations~\ref{fhthopart}. This is deduced by observing that %%@
$a\inn \ooo$ of equation~\ref{suthuoa} has 8 real components while a set of four Weyl %%@
spinors requires a total of 16 real components, or alternatively by noting that
the decomposition of the form $X =  \thX\thX^{\dag}$, as suggested in the left-hand %%@
side of equation~\ref{fhthopart}, involves an augmentation from 10 to 16 real %%@
components.
 With a complete generation of Standard Model fermions then accounted for the second %%@
and third generations might also be  directly incorporated under a further %%@
augmentation from 80 to $\sim\,$240 real components.

  Given the progression to larger symmetry groups summarised in table~\ref{lvftolvfs} %%@
from a mathematical point of view it is also natural to consider whether the Lie group %%@
$\ee$, as the largest exceptional Lie group, represented on a quintic homogeneous form %%@
$\lv$, may  mark one further and final possible step in this sequence.
 (While we refer to such a hypothetical `quintic' form, essentially an order greater %%@
than quartic is implied).
 With the smallest non-trivial representation of $\ee$ being 248-dimensional, this %%@
possibility is particularly worth consideration in light of the observations of the %%@
previous paragraph. 
  In a similar way that extending the symmetry from $\esi$ to $\ese$ led to the %%@
incorporation of right-handed as well as left-handed fermion states,  
     ideally a further extension to $\ee$ would subsume both the $\ese$ symmetry of %%@
the structure in equation~\ref{fhthopart} and explicitly  incorporate also the %%@
$u$-quark  and $\nu-$lepton spinor states and a full three generations of fermions
 all under a higher-dimensional form of $\lvh$ with an $\ee$ symmetry.

  The smallest non-trivial representation  of $\esi$ is the \textbf{27} which can be %%@
expressed as the symmetry of the cubic form $\lvt$, while for $\ese$ the  \textbf{56} %%@
representation,
 again the lowest-dimensional non-trivial representation, can be expressed as the %%@
symmetry of the quartic form $\lvfs$. However the \textbf{248} representation for 
  $\ee$ is expressed in terms of the adjoint representation on the 248-dimensional %%@
$\ee$ Lie algebra itself, with no clear interpretation in terms of a symmetry of a %%@
form of temporal flow $\lv$.
 Indeed the Lie algebra $\ee$ can be essentially introduced in terms of its action on %%@
itself,  and constructed in purely \textit{algebraic} terms which may involve the %%@
octonions \cite{Baez1},
  with the absence of any \textit{geometric} motivation or application which might be %%@
related to a homogeneous  form $\lv$.

  The fact that the smallest non-trivial representation of the 248-dimensional $\ee$ %%@
Lie algebra is expressed as the adjoint representation does not itself preclude the %%@
possibility that a \textbf{248} representation may \textit{also} be identified in %%@
terms of the symmetries of a quintic form $\lvtfe$. 
 Given the progression from the cubic polynomial form $\mbox{det}(\mcX)$ 
  of equations~\ref{detx3} and \ref{detpmn} as an expression of $\lvt$ with an $\esi$ %%@
symmetry to the terms of the quartic form $q(x)$ of equation~\ref{fquartic} underlying %%@
the form $\lvfs$ with an $\ese$ symmetry, a possible quintic form for $\lvtfe$ with an %%@
$\ee$ symmetry may be a considerably more complicated mathematical object still. Hence %%@
it is perhaps conceivable that such a structure has not been identified through purely %%@
algebraic means, even over fifty years after the corresponding $\esi$ and $\ese$ %%@
structures were first realised~\cite{Chev,Freud}.
 On the other hand if such a mathematical structure does exist, namely a quintic form %%@
$\lvtfe$ with an $\ee$ symmetry containing the form $L(\bv_4)$ with $\sltc$ symmetry, %%@
then as for the other forms of table~\ref{lvftolvfs} it \textit{would} naturally apply %%@
for the present theory, based on multi-dimensional forms of temporal flow, and further %%@
physical consequences would be \textit{expected} to be uncovered in this further %%@
progression.

In reference \cite{Asch}, as an example of a more geometrical approach, all of the %%@
classical Lie groups are accounted for as isometry groups of  bilinear or sesquilinear %%@
forms and the first four exceptional Lie groups, $\gt$, $\ff$, $\esi$ and $\ese$, are %%@
described as isometry groups constructed for cubic or quartic forms, but with $\ee$ %%@
essentially absent from the discussion. More generally little reference has been %%@
identified in the literature in which a 248-dimensional representation of $\ee$ is %%@
described in terms of an action on a quintic, or any other homogeneous polynomial, %%@
form. However in \cite{CedP, GarG}
  a polynomial of degree \textit{eight} which is invariant as a 248-dimensional %%@
representation of the compact real form of $\ee$ is described, and is closely related %%@
to an invariant polynomial for the real form $\mbox{E}_{8(8)}$.

   Considering the possible real forms of $\ee$ more generally, a suitable candidate %%@
would be 
  $\eeg$ since the following maximal subgroups involving the exceptional Lie groups %%@
are well known (see for example~\cite{Ramo}):
\begin{equation}
  \begin{array}{rcl}
   \eseg \times \mbox{SU}(1,1) & \subset & \eeg   \\
   \esig \times \mbox{SO}(1,1) & \subset & \eseg  
   \end{array}  \label{esixtoee}
\end{equation}
   This suggests the employment of the chain of non-compact real forms
   $\esig \to \eseg \to \eeg$ as symmetry groups for the corresponding forms of the %%@
sequence
   $\lvt \to \lvfs \to \lvtfe$, where the first two stages have been described here in %%@
sections~\ref{esihtho}--\ref{secesef}, while we are led to the third form as a %%@
mathematical prediction of the theory. As for the structure of the first two stages it %%@
seems likely that a construction of the final form in this progression will involve %%@
the algebraic structure of the octonions in a significant way.

  The Lie group generated by the rank-8 $\ee$ algebra, which is also uniquely the %%@
largest exceptional Lie algebra, is large enough to contain a rank-8 decomposition of %%@
the form:
\begin{equation}
 \label{eedecomp}
   \sltc \times \suth \times \sutw \times \sutw \times \uo \times \uo \subset \ee
\end{equation}
  as can be shown by straightforward  analysis of the Dynkin diagrams involved.
 Hence while the degrees of freedom of the components of $\bv_{248}$, as an extension %%@
from $\bv_{56} \equiv x \inn 
  F(\htho)$ of equation~\ref{fhthopart},  are sufficient to contain a full three %%@
generations of Standard Model fermions and a vector-Higgs, the $\ee$ symmetry group is %%@
comfortably large enough to describe the external Lorentz symmetry together with the %%@
internal $\SML$
 gauge group.

 It would be possible to attempt to identify the structures of the Standard Model, as %%@
alluded to above, within the components of a quintic form $\lvtfe$ with an $\ee$ %%@
symmetry \textit{if} the latter structure was already known and described in the %%@
literature. This would continue the approach adopted for  the $\esi$ symmetry of %%@
$\lvt$ and $\ese$ on $\lvfs$, as based on the corresponding mathematical structures %%@
originally discovered in the 1950s \cite{Chev} and 1960s \cite{Freud} respectively, %%@
for which the consequences of symmetry breaking over $M_4$ have been studied here in %%@
sections~\ref{chapesb} and \ref{secesef}.

  Alternatively the mathematical structure of $\ee$ acting on a quintic form %%@
underlying $\lvtfe$, if it exists, might itself be \textit{constructed}  
 through its application in the present theory as a form of temporal flow based on a %%@
knowledge of the empirical properties of the Standard Model.
 That is, continuing the progression of table~\ref{lvftolvfs} through the Standard %%@
Model structure identified in the components of $F(\htho)$ under the broken $\ese$ %%@
symmetry in equation~\ref{fhthopart}, and \textit{using} the need to identify spinor %%@
components for the $\nu$-lepton  and $u$-quarks, together with three generations of %%@
fermions oriented under an
 $\sutw_L$ action and relating to CKM mixing, all in a structural correspondence with %%@
the Standard Model, might lead to the identification of a suitable underlying %%@
248-dimensional space. The study of this mathematical structure, incorporating the %%@
subspaces of $\htho$ and $F(\htho)$ under the subgroups $\esi$ and $\ese$ %%@
respectively, may lead to the identification of an $\ee$ symmetry represented on the %%@
form $\lvtfe$, which might then be rigorously studied as a mathematical object in its %%@
own right, in principle with further consequences for the present theory in turn %%@
derived.

\section{Conclusions}
\label{conc}

   From the simple starting point of the multi-dimensional form of time, constrained %%@
to the form of homogeneous polynomials in the limit of infinitesimal intervals as %%@
derived for equation~\ref{lv}, the progression of forms listed in %%@
table~\ref{lvftolvfs} has been considered. 
These forms are \textit{not} however restricted to a quadratic  spacetime structure, %%@
and this has opened up the possibility to explore structures that naturally %%@
accommodate a series of Standard Model properties.

  The initial extension from a quadratic 4-dimensional spacetime form to a cubic %%@
9-dimensional temporal form, based on elements of $\hthc$, led immediately to the %%@
identification of a left-handed Weyl spinor $\psi_L$ transforming under an internal %%@
$\uo$ symmetry as described for equation~\ref{vinhinc}. On further generalisation to %%@
an $\esi$ symmetry of the 27-dimensional space $\htho$ a set of four left-handed %%@
spinors under the external Lorentz symmetry was identified, equation~\ref{thcth234}, %%@
which transform under an internal $\suth_c \times \uo_Q$ symmetry as an electron %%@
colour singlet and a $d$-quark colour triplet with the correct relative fractional %%@
electromagnetic charge arising directly out of the mathematical structure as described %%@
leading up to equation~\ref{suthuoth}. A neutrino and $u$-quark triplet are also %%@
identified provisionally in equation~\ref{suthuoa} via the properties of further %%@
components of $\htho$ under the internal $\suth_c \times \uo_Q$ transformations. 
 Structures which closely parallel $\sutw_L \times \uo_Y$ electroweak symmetry %%@
breaking were also described in section~\ref{chapesb}.
 Further extension to an $\ese$ symmetry on the 56-dimensional space $F(\htho)$ led to %%@
equation~\ref{fhthopart} with further features of the Standard Model identified, such %%@
as the suppression of the right-handed neutrino state and a left-right asymmetry more %%@
generally.

  All of the above properties have arisen out of the natural mathematical progression %%@
of the theory, based on the generalisation from spacetime symmetries as originally %%@
introduced for equation~\ref{dsxxx}, through the structures summarised in %%@
table~\ref{lvftolvfs}. At present the main prediction of the theory regards  the %%@
structure of a currently hypothetical $\ee$ symmetry action on a quintic or higher %%@
polynomial form $\lvtfe$ to complete both the progression through the exceptional Lie %%@
algebras and the full Standard Model picture.
  In the meantime the series of Standard Model properties already uncovered makes a %%@
strong case for the plausibility of the theory.
  This theory has also been explored in the directions of Kaluza-Klein geometry, %%@
quantisation scheme and applications in cosmology. These developments, together with a %%@
broader account of the work presented here, will be detailed in a separate paper, with %%@
the ultimate aim of deriving physical predictions both in high energy physics and %%@
cosmology.

%\pagebreak
%\appendix 

%\pagebreak

%Bibliography

\pagebreak

\appendix
 \section{Elements of the $\esi$ Lie Algebra}
 
\def\var{-12pt}
\begin{table}[htbp]
{\scriptsize
\centering
 %\hspace*{-18pt}
 \hspace*{3pt}
\begin{tabular}{|ccc|}
 \hline  
% boosts
   $\dot{B}_{t \pqg z}^{1}$ & $ \dot{B}_{t \pqg x}^{1}$ & $\dot{B}_{t \pqg q}^{1}$   %%@
\\
%  \vspace{\var}
  $\!\!\!\left(\!\! \begin{array}{ccc}
                    +p    &    0       &   +\fh c     \\
	                0    &   -m       &   -\fh \bar{b}      \\
                 +\fh \bar{c} &   -\fh b   &   0 
 \end{array} \!\!\right)\!\!\!$  & 
$\!\!\!\left(\!\! \begin{array}{ccc}
                  +a_x    &   \fh(p\!+\!m)       &    + \fh\bar{b}     \\
	         \fh(p\!+\!m)    &   +a_x        &   +\fh c             \\
                +\fh b   &   +\fh\bar{c} &   0
 \end{array} \!\!\right)\!\!\!$  &   
$\!\!\!\left(\!\!\!\! \begin{array}{ccc}
                  -a_q    &   \fh (p\!+\!m)q        &    +\fh q\bar{b}     \\
	        -\fh(p\!+\!m)q    &   -a_q        &  -\fh qc             \\
                -\fh bq    &   +\fh\bar{c}q  &    0
 \end{array} \!\!\!\right)\!\!\!$
    \\ \multicolumn{3}{|c|}{ }    \\  
	$\dot{B}_{t \pqg z}^{2}$ & $\dot{B}_{t \pqg x}^{2}$ & $\dot{B}_{t \pqg q}^{2}$   %%@
\\
%	\vspace{\var}
$\!\!\!\left(\!\! \begin{array}{ccc}
                    0   &  +\fh \bar{a}     &    -\fh c     \\
	                +\fh a    &    +m       &    0      \\
                    -\fh \bar{c} &    0   &    -n
 \end{array} \!\!\right)\!\!\!$ 	&   
$\!\!\!\left(\!\! \begin{array}{ccc}
                    0    &   +\fh c        &    +\fh\bar{a}     \\
	         +\fh\bar{c}  &   +b_x         &    \fh (m\!+\!n)      \\
                 +\fh a  &   \fh(m\!+\!n)    &    +b_x
 \end{array} \!\!\right)\!\!\!$  & 					
 $\!\!\!\left(\!\!\!\! \begin{array}{ccc}
                    0     &  -\fh cq       &   + \fh\bar{a}q     \\
	         +\fh q\bar{c} &  - b_q         & \fh(m\!+\!n)q           \\
                -\fh qa   &   -\fh(m\!+\!n)q    &   - b_q
 \end{array} \!\!\!\!\right)\!\!\!$  	
    \\    \multicolumn{3}{|c|}{ }   \\	 \cline{1-1} 
	 \multicolumn{1}{|c|}{ $\dot{B}_{t \pqg z}^{3}$ }               
  &  \multicolumn{1}{c}{  $\dot{B}_{t \pqg x}^{3}$ }
  &  \multicolumn{1}{c|}{ $\dot{B}_{t \pqg q}^{3}$ }   \\ 
     \multicolumn{1}{|c|}{  
           $\!\!\!\left(\!\! \begin{array}{ccc}
                   -p    &   -\fh \bar{a}        &    0     \\
	           -\fh a    &    0           &    +\fh \bar{b}      \\
                   0    &   +\fh b        &    +n
 \end{array} \!\!\right)\!\!\!$    }      &
      \multicolumn{1}{c}{  
$\!\!\!\left(\!\! \begin{array}{ccc}
                  +c_x    &   +\fh b       &    \fh(n\!+\!p)     \\
	         +\fh\bar{b}  &   0          &    +\fh a         \\
                  \fh(n\!+\!p)   &   +\fh\bar{a} &    +c_x
 \end{array} \!\!\right)\!\!\!$     } &
      \multicolumn{1}{c|}{
$\!\!\!\left(\!\! \begin{array}{ccc}
                  -c_q    &  -\fh qb          &    -\fh(n\!+\!p)q     \\
	       + \fh\bar{b}q  &   0             &    -\fh aq         \\
               \fh(n\!+\!p)q   &   +\fh q\bar{a} &    -c_q
 \end{array} \!\!\right)\!\!\!$  }
  \\
 \hline
% and rotations 
   $\dot{R}_{x \pqg q}^{1}$ & $ \dot{R}_{x \pqg z}^{1}$ & $\dot{R}_{z \pqg q}^{1}$   %%@
\\
  %   \vspace{\var}
$\!\!\!\left(\!\!\!\!\! \begin{array}{ccc}
                    0    &    -a_q\! - \! a_x q     &    -\fh qc     \\
	       -a_q\! + \! a_x q   &     0       &  +\fh q\bar{b}           \\
                  +\fh \bar{c}q  &   -\fh bq    &    0
 \end{array} \!\!\!\!\right)\!\!\!$  & 
$\!\!\!\left(\!\!\!\!\! \begin{array}{ccc}
                  +a_x    &  -\fh (p\!-\!m)       & + \fh\bar{b}     \\
	        -\fh(p\!-\!m)    &  -a_x        &    -\fh c            \\
                 +\fh b    &  -\fh\bar{c} &    0
 \end{array} \!\!\!\right)\!\!\!$  &
$\!\!\!\left(\!\!\!\!\! \begin{array}{ccc}
                 +a_q    &    \fh(p\!-\!m)q        &    -\fh q\bar{b}     \\
	        -\fh(p\!-\!m)q    &    -a_q         &    -\fh qc           \\
                 +\fh bq   &   +\fh\bar{c}q &    0
 \end{array} \!\!\!\right)\!\!\!$
    \\ \multicolumn{3}{|c|}{ }    \\   \cline{1-1} 
    \multicolumn{1}{|c|}{ $\dot{R}_{x \pqg q}^{2}$ }               
  & \multicolumn{1}{c}{  $\dot{R}_{x \pqg z}^{2}$ }
  & \multicolumn{1}{c|}{ $\dot{R}_{z \pqg q}^{2}$ }   \\
	%	\vspace{\var}
	\multicolumn{1}{|c|}{ 	
     $\!\!\!\left(\!\!\!\! \begin{array}{ccc}
                    0    &    +\fh\bar{a}q     &  -  \fh cq     \\
	           - \fh qa   &     0       & -b_q \!-\! b_xq            \\
                +\fh q\bar{c}  &  - b_q \!+\! b_xq    &    0
 \end{array} \!\!\!\!\!\right)\!\!\!$  } 	&
    \multicolumn{1}{c}{ 
$\!\!\!\left(\!\!\!\! \begin{array}{ccc}
                    0    &   +\fh c       &    -\fh\bar{a}     \\
	         +\fh\bar{c}  &   + b_x        &    -\fh(m\!-\!n)           \\
                  -\fh a  &   -\fh(m\!-\!n)    &    -b_x
 \end{array} \!\!\!\!\!\right)\!\!\!$ } &
    \multicolumn{1}{c|}{  					
 $\!\!\!\left(\!\!\!\! \begin{array}{ccc}
                    0     &   +\fh cq      &    +\fh\bar{a}q     \\
	        - \fh q\bar{c} &   +b_q         &   \fh(m\!-\!n)q       \\
                -\fh qa    &   -\fh(m\!-\!n)q    &    -b_q
 \end{array} \!\!\!\!\!\right)\!\!\!$ } 	
    \\ \multicolumn{1}{|c|}{ } & \multicolumn{2}{c|}{ }    \\	  
     \multicolumn{1}{|c|}{   $\dot{R}_{x \pqg q}^{3}$  }            
   & \multicolumn{1}{c}{  $\dot{R}_{x \pqg z}^{3}$ } 
   & \multicolumn{1}{c|}{  $\dot{R}_{z \pqg q}^{3}$  }  \\
	%  \vspace{0pt}	    
   \multicolumn{1}{|c|}{  
  $\!\!\!\left(\!\!\!\! \begin{array}{ccc}
                    0    &  +\fh q\bar{a}     &  - c_q \!+\! c_xq     \\
	            -\fh aq   &     0       &  +\fh \bar{b}q          \\
               - c_q \!-\! c_xq     &   -\fh qb    &    0
 \end{array} \!\!\!\!\right)\!\!\!$ }  &
   \multicolumn{1}{c}{  
$\!\!\!\left(\!\!\!\! \begin{array}{ccc}
                 -c_x    &   -\fh b      &    -\fh(n\!-\!p)     \\
	        -\fh\bar{b}  &   0           &    +\fh a         \\
                 -\fh(n\!-\!p)   &   +\fh\bar{a} &   +c_x
 \end{array} \!\!\!\!\!\right)\!\!\!$ }  &
    \multicolumn{1}{c|}{ 
$\!\!\!\left(\!\!\! \begin{array}{ccc}
                  -c_q    &   -\fh qb          &    -\fh(n\!-\!p)q     \\
	     + \fh\bar{b}q   &   0             &   +\fh aq         \\
               \fh(n\!-\!p)q  &   -\fh q\bar{a}  &    +c_q
 \end{array} \!\!\!\!\right)\!\!\!$  } 
   \\
 \hline
  \end{tabular}
}
  \caption{\setb    Vector fields on $T\htho$ generated by the 26 Category 1 Boosts %%@
and 31 Category 2 Rotations from table~\ref{prefbas} in the form of %%@
equation~\ref{ththo} ($\dot{B}_{t \pqg z}^{3},\dot{R}_{x \pqg q}^{2},\dot{R}_{x \pqg %%@
q}^{3} $ are non-basis elements).}
\label{lbrota}
\end{table}

\begin{table}[htbp]
\centering
\begin{tabular}{|l|}
 \hline   
$\begin{array}{lrrrrrrrr}
  \dot{A}_i:     & \; \dot{a} = &       & -a_4j & +a_3k & +a_6\kl & -a_5\jl &         %%@
&     \\
  \dot{A}_j:     & \; \dot{a} = & +a_4i &       & -a_2k & -a_7\kl &         & +a_5\il %%@
&     \\ 
  \dot{A}_k:     & \; \dot{a} = & -a_3i & +a_2j &       &         & +a_7\jl & -a_6\il %%@
&     \\
  \dot{A}_{\kl}: & \; \dot{a} = & +a_6i & +a_7j &       &         & -a_2\jl & -a_3\il %%@
&     \\
  \dot{A}_{\jl}: & \; \dot{a} = & -a_5i &       & -a_7k & +a_2\kl &         & +a_4\il %%@
&     \\
  \dot{A}_{\il}: & \; \dot{a} = &       & +a_5j & +a_6k & -a_3\kl & -a_4\jl &         %%@
&     \\
  \dot{A}_l:     & \; \dot{a} = & +a_7i & -a_6j &       &         & +a_3\jl & -a_2\il %%@
&     \\   
  \\  
  \dot{G}_i:     & \; \dot{a} = &       & -a_4j & +a_3k & -a_6\kl & +a_5\jl &-2a_8\il %%@
&+2a_7l  \\
  \dot{G}_j:     & \; \dot{a} = & +a_4i &       & -a_2k & +a_7\kl &-2a_8\jl & -a_5\il %%@
&+2a_6l  \\
  \dot{G}_k:     & \; \dot{a} = & -a_3i & +a_2j &       &-2a_8\kl & -a_7\jl & +a_6\il %%@
&+2a_5l  \\
  \dot{G}_{\kl}: & \; \dot{a} = & +a_6i & -a_7j &+2a_8k &         & -a_2\jl & +a_3\il %%@
&-2a_4l  \\
  \dot{G}_{\jl}: & \; \dot{a} = & -a_5i &+2a_8j & +a_7k & +a_2\kl &         & -a_4\il %%@
&-2a_3l  \\
  \dot{G}_{\il}: & \; \dot{a} = &+2a_8i & +a_5j & -a_6k & -a_3\kl & +a_4\jl &         %%@
&-2a_2l  \\
  \dot{G}_l:     & \; \dot{a} = & +a_7i & +a_6j &-2a_5k &+2a_4\kl & -a_3\jl & -a_2\il %%@
&        \\
  \\  \end{array} $   \\  
 $\;\,\dot{S}_q^{1}:  
  \left\{\! \begin{array}{ll}
          \dot{a} = & q\sigsum a_r r  \\
		  \dot{b} = & +\frac{3}{2}b_q - \frac{3}{2}b_1q - \fh q \sigsum b_r r \\
		  \dot{c} = & -\frac{3}{2}c_q + \frac{3}{2}c_1q - \fh q \sigsum c_r r
                         \end{array} \right. $ 
\begin{comment}  
                   & \; \dot{a} = & \multicolumn{4}{l}{q\sigsum a_r r} & 
				  \multicolumn{3}{l}{\!\!\mbox{({\color{mygray}Non-basis vectors: })}}    %%@
\\ 
				   \vspace{-20pt}  \\
  \dot{S}_q^{1}: & \; \dot{b} = & \multicolumn{7}{l}{+\frac{3}{2}b_q - \frac{3}{2}b_1q %%@
+ 
  \fh \sigsum b_r rq
              \qquad\;
   \mbox{{\color{mygray}  $ \dot{S}_q^{2}\! :  \dot{b} = 
                     q\sigsum b_r r,\;\; \dot{c}= +\ldots \!\!\!\!\!$ }}    }     \\          
       \vspace{-20pt}  \\
                  & \; \dot{c} = & \multicolumn{7}{l}{-\frac{3}{2}c_q + %%@
\frac{3}{2}c_1q +			     
	   \fh \sigsum c_r rq 
	           \qquad\; 
	\mbox{{\color{mygray}  $ \dot{S}_q^{3}\! :  \dot{c} =
                     q\sigsum c_r r,\;\; \dot{a}= +\ldots \!\!\!\!\!$ }}    }     \\  
					      \vspace{-20pt}  \\
\end{comment} 						   
  \\   \\
\hline 
 \vspace{-15pt} 
 \\
 {\footnotesize
 $\dot{S}_q^{2}:   \dot{a} = -\frac{3}{2}a_q + \frac{3}{2}a_1q - \fh q \sum a_r r, %%@
\quad\;\,
      \dot{b} =   q\sum b_r r, \quad\;\,
	     \dot{c} = +\frac{3}{2}c_q - \frac{3}{2}c_1q - \fh q \sum c_r r$   }
  \\
 {\footnotesize 
 $\dot{S}_q^{3}:  \dot{a} = +\frac{3}{2}a_q - \frac{3}{2}a_1q - \fh q \sum a_r r, %%@
\quad\;\,
     \dot{b} = -\frac{3}{2}b_q + \frac{3}{2}b_1q - \fh q \sum b_r r, \quad\;\,
	   \dot{c} = q\sum c_r r 
  $    }
 \vspace{-15pt} \\
 \\
\hline  
  \end{tabular}
 % \vspace{20pt}
  \caption{\setb Vector fields on $T\htho$ generated by the 21 Category 3 Transverse %%@
Rotations from the lower section of table~\ref{prefbas}. In the case of $\dot{A}_q$ %%@
and $\dot{G}_q$ the form of $\dot{b}=f(b)$ and $\dot{c}=f(c)$ is identical to %%@
$\dot{a}=f(a)$. With reference to equation~\ref{ththo}, in all cases $\dot{p} = %%@
\dot{m} = \dot{n} = 0$ with $\{\dot{\bar{a}}, \dot{\bar{b}}, \dot{\bar{c}}\}$  implied %%@
from $\{\dot{a}, \dot{b}, \dot{c}\}$.  ($\dot{S}_q^{2}$ and $\dot{S}_q^{3}$ are %%@
non-basis elements,
  with $\sum := \sum_{r \ne 1,q}$ here). }
\label{ltrota}
\end{table}

\par}% \linespread{1.0} for main text (28/11/15)
%match '{\setlength{\baselineskip}{0.625\baselineskip}' above

\end{document}